\documentclass[10pt,a4paper,twocolumn]{article}
\usepackage{graphicx}
\def\la{\buildrel<\over\sim}
\def\ga{\buildrel>\over\sim}
\title{\bf Fundamental Parameters of He-Weak and He-Strong Stars
\thanks{Observations taken at CASLEO, operating under agreement of CONICET and 
the Universities of La Plata, C\'ordoba and San Juan, Argentina}}

\author{L. S. Cidale $^1$, M. L. Arias$^1$, A. F.Torres$^1$, J. Zorec$^2$, Y. Fr\'emat$^3$, A. Cruzado$^1$} 
\date{}

\begin{document}
\maketitle

{$^1$Facultad de Ciencias Astron\'omicas y Geof\'\i sicas, Universidad Nacional de La Plata and Instituto de Astrof\'\i sica de La Plata (CONICET), Paseo del Bosque S/N, 1900 La Plata, Buenos Aires, Argentina} 

{$^2$Institut d'Astrophysique de Paris, UMR7095 CNRS, Universit\'e Pierre \& Marie Curie, 98bis bd. Arago, 75014 Paris, France}

{$^3$Royal Observatory of Belgium, 3 Av. Circulaire, 1180 Bruxelles, Belgium}

\abstract
{ He-weak and He-strong stars are chemically peculiar AB objects whose He lines are anomalously weak or strong for their MK spectral type. The determination of fundamental parameters for these stars is often more complex than for normal stars due to their abundance anomalies.}

{We discuss the determination of fundamental parameters: effective temperature, 
surface gravity, and visual and bolometric absolute magnitudes of He-weak and 
He-strong stars. We compare our values with those derived independently from 
methods based on photometry and model fitting.}

{We carried out low resolution spectroscopic observations in the wavelength 
range 3400-4700 \AA\ of 20 He-weak and 8 He-strong stars to determine their 
fundamental parameters by means of the Divan-Chalonge-Barbier (BCD) spectrophotometric system. This 
system is based on the measurement of the continuum energy distribution around 
the Balmer discontinuity (BD). For a few He-weak stars we also estimate the 
effective temperatures and the angular diameters by integrating absolute fluxes observed over a wide spectral range. 
Non-LTE model calculations are carried out to study the influence of the He/H 
abundance ratio on the emergent radiation of He-strong stars and on their
$T_{\rm eff}$ determination.} 

{We find that the effective temperatures, surface gravities and bolometric
absolute magnitudes of He-weak stars estimated with the BCD system and the 
integrated flux method are in good agreement between each other, and they also agree with previous
determinations based on several different methods. The mean discrepancy 
between the visual absolute magnitudes derived using the Hipparcos 
parallaxes and the BCD values is on average $\pm$ 0.3 mag for He-weak stars, while
it is $\pm$ 0.5 mag for He-strong stars. For He-strong stars, we note that the BCD calibration, based on stars in the solar environment, leads to overestimated values of $T_{\rm eff}$. By means of model atmosphere calculations  
with enhanced He/H abundance ratios we show that larger He/H ratios 
produce smaller BD which naturally explains the $T_{\rm eff}$ overestimation.
We take advantage of these calculations to introduce a method to estimate the 
He/H abundance ratio in He-strong stars. The BD of HD 37479 suggests that the 
$T_{\rm eff}$ of this star remains fairly constant as the star spectrum  undergoes changes in the intensity of H and He absorption lines. Data for the He-strong star HD 66765 are reported for the first time.}

\section{ Introduction}

 Ap/Bp stars are upper main sequence objects that show an abnormal abundance
enhancement of some chemical species in their atmospheres as compared to 
``normal" A and B dwarfs of the same effective temperature (Jaschek \& Jaschek 
\cite{JJ87a}). Besides these chemically peculiar (CP) stars, which encompass 
classical A magnetic stars (CP1), Si, Cr and SrCrEu Ap stars (CP2) and HgMn 
stars (CP3), there are also CP B-type objects having abnormally weak or 
abnormally strong lines of He\,{\sc i}. They are called He-weak (CP4) and 
He-strong stars, respectively. The He-strong stars are considered the 
high-temperature extension of classical Ap/Bp stars (Osmer \& Peterson 
\cite{OP74}).\par 
 Most CP stars have strong and variable dipolar or quadrupolar magnetic fields,
whose interaction with the gravitational and radiative diffusion processes is 
assumed to induce non-homogeneous distributions of different chemical elements.
However, in some cases, this mechanism alone cannot account for all abundance 
and isotopic anomalies (Proffitt et al. \cite{P99}).\par
 Accurate values of fundamental parameters (effective temperature, surface 
gravity, visual and bolometric absolute magnitudes) are of crucial importance 
to study the evolutionary status of  stars and the physical processes that take 
place in their atmospheres/interiors. They also enable us to explore the 
appearance of their anomalies in connection with the physical characteristics
of their local galactic environment (kinematics, magnetic fields, average 
content of chemical elements, etc.). In particular, these parameters are useful
to study the evolution of magnetic fields in CP stars.\par
 Since the photospheric abundance anomalies modify the flux distribution, the 
colours of CP stars can be affected by the concomitant blanketing effect. The
use of standard me\-thods based on calibrations of colour indices in terms of 
$T_{\rm eff}$ can then lead to erroneous results. Several authors have already
discussed the difficulties of determining the effective temperature of these 
stars (e.g. Leckrone et al. \cite{L74}; Hauck \& North \cite{HN93}; Napiwotzki 
et al. \cite{NSW93}; Sokolov \cite{S98}). From these studies it is apparent 
that the use of the Geneva or Str\"omgren photometry to determine $T_{\rm 
eff}$ of CP stars is much more complex than for normal stars. Napiwotski et al. 
(\cite{NSW93}) have discussed different uvby$\beta$ photometric calibrations 
to derive  $T_{\rm eff}$ and $\log g$. They recommended the use of Moon \& 
Dworetsky's (1985) calibration corrected for gravity deviations.\par
 Other alternative methods have also been proposed. For instance, the infrared flux method (IRFM) which minimizes the blanketing effect (Blackwell \& Shallis \cite{BS77}),  measures the ratio of the 
total integrated flux to a monochromatic IR flux. This method is often used to determine angular diameters 
and effective temperatures of normal as well as of peculiar stars. Very 
frequently, the fundamental parameters are also determined by fitting 
synthetic spectra to the observed absolute energy distributions, or to line
profiles, particularly the H$\beta$ line. For CP2 stars, which are generally
too blue for their spectral type, a supplementary correction to the effective 
temperature is needed (Stepie\'n \& Dominiczak \cite{SD89}).\par 
 Sokolov (\cite{S98}) proposed the Balmer continuum slope near the Balmer jump 
as a tool to determine the $T_{\rm eff}$ of CP stars. The method takes 
advantage of the fact that this slope is quite similar for normal and 
chemically peculiar main sequence stars. Although this method is easy to apply,  it requires the stellar spectrophoto\-me\-try to be corrected both
for ISM reddening and for possible effects due to the abnormal energy 
distribution of CP stars.\par
 The surface gravity parameter $\log g$ is often determined through theoretical line profile fittings  or by using evolutionary tracks. To
enter the evolutionary tracks the absolute bolometric magnitude $M_{\rm bol}$
and $T_{\rm eff}$ of stars are needed. Most frequently, $M_{\rm bol}$ is 
determined with photometric data and trigonometric parallaxes which provide 
the visual absolute magnitude that, in turn, is transformed into $M_{\rm bol}$ 
with the help of a bolometric correction calibration (North et al. \cite{N97}; 
G\'omez et al. \cite{G98}, Hubrig et al.\cite{H04}, Kochukhov \& Bagnulo 
\cite{KB06}).\par
  Not only are these methods difficult to apply to CP stars but also none of them give simultaneously $T_{\rm eff}$, $\log g$, $M_{\rm V}$ and $M_{\rm bol}$. Particularly, a) the photometry-based methods need to take into account 
line blanketing effects, or they rely on corrections from chemical anomalies 
that are applied to colours; b) the measurement of the 
Balmer continuum slope is based on data previously corrected for energy redistributions and/or reddening effects; and c) the determination of visual absolute 
magnitudes from where the bolometric magnitudes are obtained, needs 
well-determined parallaxes and effective temperatures.\par 
 Since most of these methods were applied to CP2 and CP3 categories, in the
present contribution we focus our attention on He-weak and He-strong type
stars. We note that information on He-strong stars is particularly scarce. In
this work, we propose the use of the BCD spectrophotometric system (Barbier \& 
Chalonge \cite{BC41}; Chalonge \& Divan \cite{CD52}) as an alternative 
tool to infer spectral types, effective temperatures, surface 
gravities, visual and bolometric absolute magnitudes of CP stars. This method has 
the advantage of being based on direct measurable quantities of the continuum 
energy distribution in wavelengths near the the Balmer discontinuity (BD), which do not need corrections 
for interstellar and/or circumstellar extinctions and cannot be altered 
for beforehand unknown chemical anomalies. These quantities are strongly 
sensitive to the ionization balance of stellar atmospheres, so that they are 
excellent indicators of $T_{\rm eff}$ and $\log g$. Since these parameters 
certainly depend somehow on the chemical properties of stars, it seems useful 
to inquire whether or not we can also draw from them quantitative indications on 
their chemical anomalies.\par
 In \S 2 we present the observational material. In \S 3 we give a summary 
about the BCD spectrophotometric system and complementary techniques used to test against the BCD system. \S 4 is devoted to the determination of  fundamental parameters of 20 He-weak stars with the BCD system. We also include our determinations of effective temperatures, angular 
diameters and bolometric magnitudes done with the integrated flux method. We discuss the  results with those obtained elsewhere independently. \S 5 gives fundamental parameters derived for 8 He-strong stars. Non-LTE model low resolution stellar spectra are also used to obtain the effective temperature and to analyze the effect of enhanced He/H abundance ratio on the BD. We discuss 
 the difficulties encountered in determining the fundamental parameters of 
He-strong stars and  the variation of HD 37479 spectrum  mainly in view of its $T_{\rm eff}$ estimate. \S 6 shows the evolutionary status of the studied He-peculiar stars. Due to the variability observed in the He-strong group we add an appendix containing line equivalent widths. Conclusions are given in \S 7.\par 

\section{Observations}

 We obtained low resolution spectra of 28 CP stars: 20 He-weak and 8 He-strong stars (see Table \ref{obs}). Most stars were observed only once. Observations were carried out at the Complejo Astron\'omico El Leoncito (CASLEO), San Juan, Argentina, with
the 2,15$\,$m telescope and the Boller \& Chivens spectrograph on August 
30-31, September 1-3 and September 28-30, 2004, and on January 18-19, 2006. We 
used a 600 l\,mm$^{-1}$ grating (\# 80), a slit width of 250 $\mu$ and a CCD 
detector of 512 $\times$ 512 pixels. The obtained spectra cover the range 
3400-4700 \AA\ with an effective resolution of 4.53 \AA\ each 2 pixels.\par 
 Bias, flat field, He-Ne-Ar comparison and spectrophotometric flux standard 
star spectra were also secured to perform wavelength and flux calibrations.  
The reduction of observations was made with the {\sc 
iraf}\footnote[1]{{\sc iraf} is distributed by the National Optical Astronomy
Observatory, which is operated by the Association of Universities for 
Research in Astronomy (AURA), Inc., under cooperative agreement with the 
National Science Foundation} software package (version 2.11.3) and all the 
spectra were corrected from atmospheric extinction.\par

\begin{table}[pt]
\caption{Program  He-weak (top) and He-strong (botton) stars. In the first two columns we give the star identifiers. Column 3 shows the date of observation. The peculiarity class, and the average quadratic effective magnetic field are presented in columns 4 and 5 and were taken from Bychkov et al. (2003)}
\label{obs}
\begin{center}
\tabcolsep 3.5pt
\begin{tabular}{rclll}
\hline\hline
\noalign{\smallskip}
~~HD & HR    & Date &  Peculiarity & $<B_{e}>$ \\
   &       & dd/mm/yy &              &  Gauss     \\
\hline
\noalign{\smallskip}
 {  5737}  & 280  & 01/09/04 & SrTi He-w  & 324.0 \\
                   &      & 03/09/04 &             &       \\
                   &      & 04/09/04 &             &       \\
 {  19400} & 939  & 01/09/04 & He-w        & 206.7 \\
 {  22470} & 1100 & 01/09/04 & Si He-w     & 732.9 \\
 {  22920} & 1121 & 30/09/04 & Si4200 He-w & 307.1 \\
                   &      & 01/10/04 &             &       \\
 { { 23408}} & 1149 & 03/09/04 & Hg He-w     & 410.0  \\ 
 { { 28843}} & 1441 & 03/09/04 & Si He-w     & 344.2  \\ 
 { { 49333}} & 2509 & 03/09/04 & Si He-w     & 618.4 \\ 
                   &      & 01/10/04 &             &       \\
 { { 49606}} & 2519 & 01/10/04 & MnHgSi He-w & 916.0 \\
 { { 51688}} & 2605 & 01/10/04 & Hg He-w     & 550.2  \\
 { { 57219}} & 2790 & 04/09/04 &             & -- \\
                   &      & 18/01/06 &             &     \\ 
 { { 74196}} & 3448 & 18/01/06 & He-w        & --\\
 { { 125823}}& 5378 & 18/01/06 & Si He-w     & 469.3 \\
 { { 142301}}& 5912 & 01/10/04 & Si He-w     & 2103.6 \\  
 { { 142990}}& 5942 & 02/09/04 & He-w        & 1304.3  \\
                   &      & 30/09/04 &             &    \\
 { { 144334}}& 5988 & 01/10/04 & Si He-w     & 783.2 \\ 
 { { 144661}}& 5998 & 01/10/04 & He-w        & 542.0  \\
 { { 144844}}& 6003 & 01/10/04 & Si He-w     & 318.1 \\ 
 { { 162374}}& 6647 & 01/10/04 & He-w        & 269.8 \\
 { { 175362}}& 7129 & 03/09/04 & SiMn He-w  & 3569.9 \\
 { { 202671}}& 8137 & 01/09/04 & SrTi He-wm  & 183.0 \\ 
\hline
\noalign{\smallskip}
{ { 37017}} & 1890 & 01/09/04 & He-s & 1488.1  \\ 
                  &      & 30/09/04 &      &                \\
{ { 37479}} & 1932 & 02/09/04 & He-s & 1907.9  \\
                  &      & 04/09/04 &      &       \\
{ { 37776}} & BD-01 1005 & 04/09/04  & He-s & 1259.7 \\
{ { 58260}} & CD-36 3578 & 18/01/06  &He-s  & 2291.2 \\
{ { 60344}} & CD-23 5673 & 18/01/06  &He-s  & 334.9 \\
{ { 64740}} & 3089 & 03/09/04  & He-s & 571.9 \\
                  &      & 30/09/04  &      & \\
{ { 66765}} & CD-47 3537 & 18/01/06   & He-s & --\\
{ {264111}} & BD+04 1447 & 18/01/06   & He-s & --\\
\noalign{\smallskip}
\hline
\end{tabular}
\end{center}
\end{table}  

\section{The Methods}

\subsection{The BCD spectrophotometric system}

 The fundamental parameters of the program stars were mainly derived by using the BCD 
spectrophotometric system. This system is based on the observable parameters 
($\lambda_1, D$). $D$ measures the Balmer jump at $\lambda$3700 \AA\ and it is a 
strong indicator of the effective temperature. $\lambda_1$ gives the mean 
spectral position of the Balmer jump and it is related to the surface 
gravity.\par

 The BCD spectrophotometric system has several advantages: 1) it is not 
affected either by interstellar or circumstellar extinction (Zorec \& Briot 
\cite{ZB91}), 2) it is based on measurable parameters of the visual continuum 
energy distribution near the BD which are thus related, on average, to
physical properties of deeper photospheric layers than any classification 
system based on spectral line measurements, 3) it is possible to derive 
information on most of the fundamental parameters, 4) this method is easy and
direct to apply, and 5) it can be applied with high accuracy to normal and 
many peculiar classes of B-type stars. Zorec \& Briot (\cite{ZB91}), Cidale et 
al. (\cite{CZ01}) and Zorec et al. (\cite{ZFC05}) applied it successfully to Be and 
B[e] objects, even though these objects may display a second BD that can be 
either in emission or in absorption (Divan \cite{D79}).\par
 To determine the parameter $D=$ $\log(F_{+3700}/F_{-3700}$) we normalize the 
observed energy distribution with a Planckian function, which rectify the 
energy distribution at both sides of the BD, and then, we extrapolate the 
Paschen continuum to obtain the flux at $F_{+3700}$. To determine $F_{-3700}$, 
we use the flux level where the higher members of Balmer lines merge together. 
The use of Balmer lines to determine the low level of the Balmer jump has the 
advantage of giving an estimate of the BD which is not affected by possible 
variations of the intensity of the Balmer continuum.\par
 We have measured $D$ and $\lambda_1$ for the program stars and determined 
their spectral type, $T_{\rm eff}$, $\log g$, and absolute magnitudes $M_{\rm 
V}$ and $M_{\rm bol}$ (see details in \S 4 and \S 5) using the respective 
calibrations of ($\lambda_1,D$) by Divan \& Zorec (\cite{dz82}), Zorec 
(\cite{z86}), and Zorec \& Briot (\cite{ZB91}). The uncertainties on $T_{\rm 
eff}$, $\log g$, $M_{\rm V}$ and $M_{\rm bol}$ are mainly related to the 
errors of the BCD observational quantities ($\lambda_1,D$). Typical values are 
$\sigma(D)\!\la\!0.015$ dex and $\sigma (\lambda_1)\!\sim\!5$\AA, which 
produce $\Delta T_{\rm eff}\!\sim\pm500$ K for the late B-type stars and 
$\pm1500$ K for the early B-types ($T_{\rm eff}\,>$ 20000 K), $\Delta\log 
g\!\sim\!\pm0.2$ dex, and $\Delta M\!\sim\pm0.3$ mag both for $M_{\rm V}$ and
$M_{\rm bol}$.\par

\subsection{The integrated-flux method}

 In addition to the BCD classification system we used the integrated-flux method to derive fundamental parameters for those He-weak stars of our program with absolute fluxes observed over a wide spectral range.  We perform a 
determination of the effective temperature, angular diameter ($\theta$), and
bolometric magnitude by means of the relation:
\begin{equation}
F_{\rm bol} = (\theta^2/4)\sigma_{\rm\check{S}B}T_{\rm eff}^4
\label{tbol}
\end{equation}
\noindent where $F_{\rm bol}$ is the absolute bolometric flux reduced to the 
distance to the Earth and corrected for interstellar extinction; $\sigma_{\rm
\check{S}B}$ is the \v Stefan-Boltzmann constant. Relation (\ref{tbol}) is
i\-te\-ra\-ted starting with an approximate $T_{\rm eff}$ until the 
difference between two succesive iterations becomes $\Delta T_{\rm eff}\!\la\! 1$ K. 
This implies from 10 to 30 iteration steps. At each step, $\theta$ is calculated 
using the $F_{\lambda}/F_{\rm model}$ ratio of fluxes in the wavelength 
interval $\lambda\lambda 7000-8000$ \AA. The $F_{\lambda}/F_{\rm model}$ 
ratios are fairly insensitive to $\log g$. The observed fluxes $F_{\lambda}$ 
are corrected for ISM extinction and the $F_{\rm model}$ corresponds to the running $T_{\rm eff}$. At each step we also determine the amount of non-observed 
far-UV and far-IR energy as a function of the $T_{\rm eff}$ under iteration
using LTE models (Kurucz \cite{kur92}), which we add to the integrated observed flux 
to obtain $F_{\rm bol}$. The wavelength interval of observed absolute fluxes 
ranges from 1300 \AA\ to 1 $\mu$m. The fluxes are observed in the far-UV by the TD1 and IUE satellites, and from visible to near-IR 
with the 13-colour narrow band photometry calibrated in absolute fluxes 
(Johnson \& Mitchell \cite{JM75}).

\subsection{Model fitting}

 In the case of He-strong stars we also determine  $T_{\rm eff}$, 
and $\log g$ by computing NLTE synthetic spectra with the TLUSTY
and SYNSPEC computer codes (Hubeny \& Lanz \cite{HL95}, and the references 
therein) assuming H-He model atmospheres and He/H ratios of 0.1, 0.2, 0.5 and 
1.0 . The atomic models we used are basically those provided on the TLUSTY 
website for H\,{\sc i} (9 levels), He\,{\sc i}  (20 individual levels) and 
He\,{\sc ii} (20 levels). The Stark broadening of the He\,{\sc i} transitions
close to the (1s2p) 3Po ionization limit is taken from Dimitrijevic \& Sahal-Brechot (\cite{DS84},\cite{DS90}), or it is computed using an approximate relation proposed by Freudenstein \& Cooper (\cite{FC78}). In all cases, a microturbulent velocity of  2 km~s$^{-1}$ is assumed. To have synthetic spectra comparable with the ones observed at CASLEO, we resampled them to the resolution of the observed energy distributions.

\section{He-weak stars}

Table~\ref{he-w1} gives measurements and fundamental parameters derived from the BCD spectrophotometric system. Table~\ref{he-w2} reproduces the values of the fundamental parameters derived by using integrated absolute fluxes.

\begin{table}[t]
\caption{He-weak stars: measured ($\lambda_1$,$D$) BCD quantities and the corresponding fundamental parameters: $T_{\rm eff}$, $\log g$, spectral type, $M_{\rm V}$, and $M_{\rm bol}$ determined using the calibrations of ($\lambda_1,D$) for OB stars of solar He/H ratio.}
\label{he-w1}
\begin{center}
\tabcolsep 4.0pt
\begin{tabular}{rccrclcc}
\hline\hline
\noalign{\smallskip}
HD & $\lambda_1$ & $D$ & $T_{\rm eff}$ & $\log g$ & \scriptsize{Sp.T} & $M_{\rm V}$ &  $M_{\rm bol}$ \\
     & \AA         & dex & K             & dex      &      & mag         &  mag \\
\hline
\noalign{\smallskip}
{ { 5737}}   & 32.74 & 0.281 & 12700 & 2.95 &\scriptsize{B6III} & $-$2.5 & $-$3.4 \\ 
{ { 19400}}  & 48.30 & 0.320 & 13000 & 3.90 &\scriptsize{B6-7V} & $-$0.8 & $-$1.5 \\
{ { 22470}}  & 58.68 & 0.296 & 14000 & 4.21 &\scriptsize{B6V}   & $-$0.3 & $-$1.6 \\
{ { 22920}}  & 46.10 & 0.277 & 14200 & 3.80 &\scriptsize{B6IV}  & $-$1.5 & $-$2.5 \\
{ { 23408}}  & 39.02 & 0.408 & 11250 & 3.20 &\scriptsize{B9III} & $-$1.1 & $-$1.6 \\
{ { 28843}}  & 50.67 & 0.273 & 14700 & 3.95 &\scriptsize{B5IV}  & $-$1.1 & $-$2.2 \\
{ { 49333}}  & 61.20 & 0.250 & 15800 & 4.23 &\scriptsize{B5V}   & $-$0.7 & $-$2.3 \\ 
{ { 49606}}  & 46.53 & 0.340 & 13500 & 3.80 &\scriptsize{B6IV}  & $-$1.2 & $-$2.0 \\
{ { 51688}}  & 49.98 & 0.340 & 12500 & 3.95 &\scriptsize{B7-8IV}& $-$0.6 & $-$1.2 \\
{ { 57219}}  & 51.91 & 0.193 & 18500 & 3.92 &\scriptsize{B2.5IV}& $-$2.0 & $-$3.7 \\
{ { 74196}}  & 54.78 & 0.299 & 13750 & 4.13 &\scriptsize{B6V}   & $-$0.5 & $-$1.6 \\
{ { 125823}} & 57.64 & 0.184 & 19400 & 4.10 &\scriptsize{B2V}   & $-$1.7 & $-$3.6 \\
{ { 142301}} & 65.60 & 0.252 & 15700 & 4.30 &\scriptsize{B5V}   & $-$0.7 & $-$2.2 \\ 
{ { 142990}} & 67.21 & 0.208 & 18500 & 4.27 &\scriptsize{B3V}   & $-$1.2 & $-$3.0 \\ 
{ { 144334}} & 63.76 & 0.280 & 14500 & 4.30 &\scriptsize{B5-6V} & $-$0.4 & $-$1.8 \\ 
{ { 144661}} & 60.31 & 0.273 & 15000 & 4.23 &\scriptsize{B5V}   & $-$0.5 & $-$2.0 \\
{ { 144844}} & 65.60 & 0.346 & 12100 & 4.32 &\scriptsize{B8V}   & $+$0.4 & $-$0.8 \\
{ { 162374}} & 51.11 & 0.247 & 15700 & 3.95 &\scriptsize{B5V}   & $-$1.3 & $-$2.7 \\
{ { 175362}} & 64.75 & 0.219 & 17500 & 4.25 &\scriptsize{B3V}   & $-$1.0 & $-$3.0 \\
{ { 202671}} & 44.37 & 0.303 & 13200 & 3.70 &\scriptsize{B6IV}  & $-$1.3 & $-$2.0 \\
\noalign{\smallskip}
\hline
\noalign{\smallskip}
\noalign{\smallskip}
\multicolumn{8}{l}{Note: The $\lambda_1$ parameter is given in $\lambda-3700$ \AA}\\
\noalign{\smallskip}
\hline 
\end{tabular}
\end{center}
\end{table}
 HD 125823 is the only star whose He\,{\sc i} lines vary in strength between
anomalously weak and anomalously strong. It was included in Table~\ref{he-w1}
because the star presented a He-weak phase at the date of our 
observations.\par
\begin{table}[ht]
\caption{He-weak stars: our determinations of fundamental parameters: $T_{\rm eff}$, $\theta$,$F_{\rm bol}$, $M_{\rm bol}$ and the $E(B-V)$ color excess used for the ISM extinction correction applying the integrated-flux method.}
\label{he-w2}
\begin{center}
\tabcolsep 2.2pt
\begin{tabular}{rlrrcc}
\hline\hline
\noalign{\smallskip}
HD & ~~~~$T_{\rm eff}$(\sc{IF}) & $\theta$(rad)~~~   & $F_{\rm bol}$~~~~   & $M_{\rm bol}$ &  $E_{B-V}$ \\
   &   ~~~~~~~ K            & $\times10^{-10}$~~ & $\times10^{-7}$~ & mag & mag \\
\hline
\noalign{\smallskip}
  5737 & $12490\pm143$ & $17.6\pm0.2$ & $10.6\pm0.6$ &-3.1$\pm0.4$ & 0.00\\
 57219 & $18286\pm125$ & $ 8.9\pm0.1$ & $12.5\pm0.5$ &-3.8$\pm0.3$ & 0.05\\
125823 & $17594\pm172$ & $12.8\pm0.2$ & $22.2\pm1.7$ &-2.9$\pm0.2$ & 0.01\\
142301 & $15084\pm~84$ & $ 8.8\pm0.1$ & $ 5.7\pm0.2$ &-1.6$\pm0.4$ & 0.12\\
144334 & $15099\pm~77$ & $ 8.1\pm0.1$ & $ 4.8\pm0.2$ &-1.6$\pm0.3$ & 0.10\\
144661 & $13421\pm121$ & $ 7.6\pm0.5$ & $ 2.7\pm0.4$ &-0.4$\pm0.2$ & 0.09\\
144844 & $12448\pm~50$ & $10.1\pm0.1$ & $ 3.5\pm0.1$ &-0.9$\pm0.2$ & 0.12\\
\noalign{\smallskip}
\hline
\end{tabular}
\end{center}
\end{table}

\subsection{$T_{\rm eff}$ determination}

In order to test the accuracy  of the BCD effective temperatures a comparison with the values obtained by other methods is performed. Table~\ref{comp} shows the effective temperatures with their corresponding uncertainties (when available) derived from the BCD system and from the literature.

\begin{table}[t]
\begin{center}
\caption{Comparison of the BCD effective temperatures with those determined by 
other authors for the program He-weak stars. Column 1 gives the HD number; 
column 2, our $T_{\rm eff}$(BCD) values; column 3, $T_{\rm eff}$(G) 
 (Glagolevskij \cite{G02}) determined from the parameters Q in the UBV
system and X in the multicolour system; column 4, $T_{\rm eff}$(KB) 
determined by Kochukhov \& Bagnulo (\cite{KB06}) using the calibration of
the Geneva photometry (Golay \cite{G72}) and the method suggested by Hauck \& 
North (\cite{HN93}); column 5: $T_{\rm eff}$(HG) from Hunger \& Groote
(\cite{HG99}) using the IRFM, column 6: $T_{\rm eff}$(L) obtained by Lanz 
(\cite{L85}) using the IRFM, column 7: $T_{\rm eff}$(HN) from Hauck \& North 
(\cite{HN93}) using the Geneva photometry, and column 8, other values of 
$T_{\rm eff}\,\pm \Delta T$ found in the literature, most of them determined by model fitting}
\scriptsize
\label{comp}
\tabcolsep 3.0pt
\begin{tabular}{rccccccc}
\hline\hline
\noalign{\smallskip}
HD & \multicolumn{7}{c}{$T_{\rm eff}\times10^3$ K} \\
\noalign{\smallskip}
\cline{2-8} 
\noalign{\smallskip}
   & BCD & G~~ & KB~ & HG~ & L~~ & HN~ & Others \\
\noalign{\smallskip}
\hline
\noalign{\smallskip}
  5737 & 12.7$\pm$0.5 & 13.50$\pm$0.3 & 13.2$\pm$0.4 & 14.4$\pm$0.1 &  --            &  --   & 13.6$\pm$0.2$^{a}$ \\
       &              &               &              &              &                &       & 14.0$^{b}$ \\
 19400 & 13.0$\pm$0.5 & 12.90$\pm$0.3 &   --         &         --   &  --            &  --   & 13.3$\pm$0.5$^{a}$ \\
       &              &               &              &              &                &       & 13.9$^{c}$ \\
 22470 & 14.0$\pm$0.5 & 13.45$\pm$0.3 & 13.0$\pm$0.4 & 13.5$\pm$0.4 & 13.79$\pm$0.14 & 14.25 & 13.4$\pm$0.5$^{a}$ \\
 22920 & 14.2$\pm$0.5 & 14.40$\pm$0.3 & 13.9$\pm$0.4 & 14.3$\pm$0.3 &   --           &  --   & 14.1$\pm$0.5$^{a}$ \\
 23408 & 11.2$\pm$0.5 & 12.30$\pm$0.3 & 12.0$\pm$0.4 &         --   &   --           & 13.80 &  --         \\
 28843 & 14.7$\pm$0.5 & 14.53$\pm$0.3 & 13.9$\pm$0.4 & 15.0$\pm$0.5 & 14.80$\pm$0.19 & 15.28 &  --         \\ 
 49333 & 15.8$\pm$0.5 & 16.60$\pm$0.3 & 16.4$\pm$0.5 & 15.7$\pm$0.4 & 15.94$\pm$0.22 & 16.81 &  --         \\
 49606 & 13.5$\pm$0.5 & 13.80$\pm$0.3 &  --          &         --   &  --            &  --   & 14.8$^{b}$ \\
 51688 & 12.5$\pm$0.5 & 13.05$\pm$0.3 &  --          &         --   &  --            &  --   &  --         \\
 57219 & 18.5$\pm$0.5 & 16.35$\pm$0.3 &  --          &         --   &  --            &  --   & 17.1$^{d}$ \\
 74196 & 13.7$\pm$0.5 & 13.30$\pm$0.3 &  --          & 14.2$\pm$0.6 &  --            &  --   &  --         \\
125823 & 19.4$\pm$0.5 & 19.53$\pm$0.3 & 17.7$\pm$0.5 & 18.4$\pm$0.4 & 19.01$\pm$0.23 & 19.60 &  --         \\
142301 & 15.7$\pm$0.5 & 16.50$\pm$0.3 & 15.6$\pm$0.4 & 15.9$\pm$0.3 & 15.99$\pm$0.15 & 17.26 &  --         \\
142990 & 18.5$\pm$0.5 & 17.80$\pm$0.3 & 16.5$\pm$0.5 & 16.9$\pm$0.5 & 18.02$\pm$0.23 &  --   &  --         \\
144334 & 14.5$\pm$0.5 & 15.40$\pm$0.3 & 14.7$\pm$0.4 & 15.2$\pm$0.5 & 14.58$\pm$0.23 & 16.25 &  --         \\
144661 & 15.0$\pm$0.5 & 15.00$\pm$0.3 &   --         &          --  &  --            &   --  &  --         \\
144844 & 12.1$\pm$0.5 & 12.30$\pm$0.3 &   --         &          --  &  --            &  --   &  --         \\
162374 & 15.7$\pm$0.5 & 17.30$\pm$0.3 &   --         & 16.1$\pm$0.3 & 17.84$\pm$0.22 & 17.37 &  --         \\
175362 & 17.5$\pm$0.5 & 17.00$\pm$0.3 & 16.5$\pm$0.4 &         --   & 16.38$\pm$0.20 & 18.24 & 16.5$^{e}$ \\
202671 & 13.2$\pm$0.5 & 14.10$\pm$0.3 &   --         &         --   &  --            &  --   & 13.1$\pm$0.2$^{a}$ \\
\noalign{\smallskip}
\hline
\noalign{\smallskip}
\multicolumn{8}{l}{$a$:  Leone \& Manfr\'e (\cite{LM97}) or Leone et 
al.(\cite{L97}), $b$: Cayrel De Strobel et al.(\cite{CHF92}) }\\
\multicolumn{8}{l}{ $c$: Jaschek \& 
Jaschek (1987b), $d$: Kroll (\cite{KR87}), $e$: Meg\'essier (1988b)} \\
\noalign{\smallskip}
\hline
\end{tabular}
\end{center}
\end{table}

 Except for few stars,  the effective 
temperatures derived with the BCD method have no significant differences with the values derived from Geneva 
and other multicolour photometric systems, the IRFM and LTE line-blanketed models. The mean discrepancies found
between our data and those derived by other authors are lower than 800 K. The
best agreement is obtained with the IRFM (Lanz \cite{L85}), the mean 
temperature difference being $|\Delta$ $T_{\rm eff}|\!\sim\!$ 500 K.
 We have also obtained a good agreement between our own determinations of $T_{\rm eff}(IF)$ by means of the integrated-flux method and the BCD system. Comparing Tables~\ref{he-w1} and \ref{he-w2}, we found a mean value of  $\Delta\!T_{\rm eff}\,=\,$$\!\pm\!700 K$.

\begin{figure}
\centering
\includegraphics[width=8cm,height=8cm,angle=270]{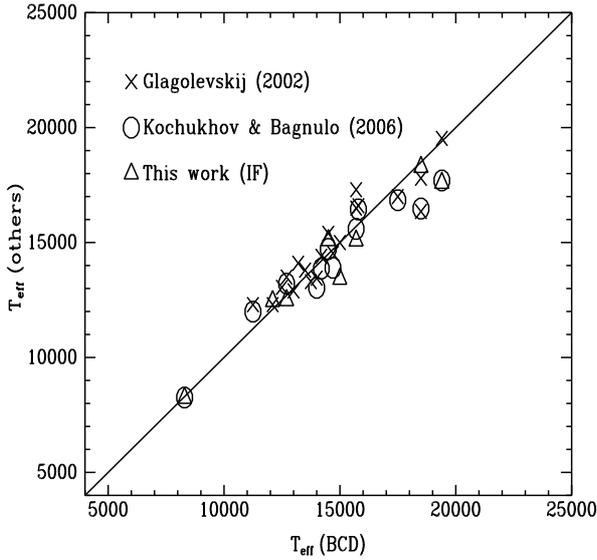}
\caption{Comparison of the effective temperatures derived from the BCD 
system with those obtained photometrically by Glagolevskij (\cite{G02}), 
Kochukhov \& Bagnulo (\cite{KB06}) and in this work with the integrated flux 
method}
\label{temp}
\end{figure}

 Figure~\ref{temp} displays the comparison of our $T_{\rm eff}$(BCD) values 
{\it vs.} $T_{\rm eff}$(G) (crosses), $T_{\rm eff}$(KB) (open circles), and  
$T_{\rm eff}$(IF) (open triangles). In the same figure, it is also plotted a reference line with slope 1. We selected the mentioned works to perform the
comparison, since the number of stars in common is large. The good correlation 
found between the BCD system and other methods confirms that the BD of He-weak 
stars behaves like that observed in normal B stars. For these stars we can 
then safely use the existent calibrations, done for He/H $\approx0.1$, to 
determine their fundamental parameters.\par 

From our results we conclude that the behaviour of the Balmer jump in He-weak stars is different from that of Ap stars (CP2 group). The photometric 
anomalies of Ap stars (Gerbaldi et al. \cite{GHM74}; Hauck \& North \cite{HN82}) may introduce noticeable differences between the BCD and MK spectral 
classifications (Chalonge \& Divan \cite{CD77}).\par

\subsection{The surface gravity}
\label{slogg}

Surface gravity determinations for He-peculiar stars are hardly common. Hunger \& Groote (1999) determined $\log\,g$ values from {\sc hipparcos} parallaxes together with tracks from stellar evolution, Leone \& Manfr\`e (1996) estimated values using synthetic H line profiles while Leone et al. (1997) used Moon \& Dworetsky's (1985) calibration and applied the corrections suggested by Napiwotzki et al (1993). Table \ref{g-hew} compares, for each object, the $\log\,g$ values determined by means of the BCD system with those derived by the above mentioned works. We find an excellent agreement with the values given by Hunger \& Groote (1999), $|\Delta \log\, g|$ $<$ 0.13 dex, for 11 stars in common and a good agreement with Leone's works, $|\Delta \log\, g|$ = 0.23 dex for six stars in common.\par

\begin{table}[ht]
\caption{He-weak stars: comparison of BCD $\log\,g$ values with those obtained by other authors}
\label{g-hew}
\begin{center}
\tabcolsep 4.0pt
\begin{tabular}{rccc}
\hline\hline
\noalign{\smallskip}
HD & $\log\!g(BCD)$  & $\log\!g(HG)$ & $\log\!g(Le)$  \\
     &   dex         & dex           & dex                \\
\hline
\noalign{\smallskip}
{ { 5737}}   &  2.95$\pm$0.2 & 3.48$_{-0.11}^{+0.15}$  & 3.20$\pm$0.10 \\ 
{ { 19400}}  &  3.90$\pm$0.2 &  --                     & 3.76$\pm$0.25 \\
{ { 22470}}  &  4.21$\pm$0.2 & 3.97$_{-0.08}^{+0.09}$  & 4.15$\pm$0.25 \\
{ { 22920}}  &  3.80$\pm$0.2 & 3.82$_{-0.11}^{+0.13}$  & 3.72$\pm$0.25 \\
{ { 23408}}  &  3.20$\pm$0.2 &  --                     & --    \\
{ { 28843}}  &  3.95$\pm$0.2 & 4.33$_{-0.08}^{+0.08}$  & 4.25$\pm$0.25\\
{ { 49333}}  &  4.23$\pm$0.2 & 4.18$_{-0.10}^{+0.12}$  & 4.08$\pm$0.25\\ 
{ { 49606}}  &  3.80$\pm$0.2 &  --                     & 3.89$\pm$0.25\\
{ { 51688}}  &  3.95$\pm$0.2 &  --                     & --  \\
{ { 57219}}  &  3.92$\pm$0.2 &  --                     & --  \\
{ { 74196}}  &  4.13$\pm$0.2 & 4.14$_{-0.05}^{+0.30}$  & --  \\
{ { 125823}} &  4.10$\pm$0.2 & 4.16$_{-0.07}^{+0.08}$  & --  \\
{ { 142301}} &  4.30$\pm$0.2 & 4.29$_{-0.11}^{+0.13}$  & --  \\ 
{ { 142990}} &  4.27$\pm$0.2 & 4.20$_{-0.09}^{+0.10}$  & --  \\ 
{ { 144334}} &  4.30$\pm$0.2 & 4.29$_{-0.09}^{+0.10}$  & --  \\ 
{ { 144661}} &  4.23$\pm$0.2 &  --                     & --  \\
{ { 144844}} &  4.32$\pm$0.2 &  --                     & --  \\
{ { 162374}} &  3.95$\pm$0.2 & 3.95$_{-0.14}^{+0.18}$  & --  \\
{ { 175362}} &  4.25$\pm$0.2 &  --                     & 3.70$\pm$0.10\\
{ { 202671}} &  3.70$\pm$0.2 &  --                     & 3.40$\pm$0.10 \\
\noalign{\smallskip}
\hline
\noalign{\smallskip}
\multicolumn{4}{l}{Column 2: this work, Column 3: Hunger \& }\\ 
\multicolumn{4}{l}{Groote (1999), Column 4: Leone \& Manfr\'e (\cite{LM97})}\\ 
\multicolumn{4}{l}{or Leone et al.(\cite{L97})}\\
\hline
\noalign{\smallskip}
\end{tabular}
\end{center}
\end{table}

\begin{table}[ht]
\begin{center}
\caption{He-weak stars: Comparison of the BCD absolute visual magnitudes with those derived by other authors. The uncertainties of the values are quoted when they are available}
\label{magni}
\begin{tabular}{rrrrr}
\hline
\hline
~~ HD &$M_{\rm V}$(BCD) & $M_{\rm V}$(G) & $M_{\rm V}$(KB) & $M_{\rm V}$(Go) \\
\noalign{\smallskip}
\hline
\noalign{\smallskip}
  5737   & -2.5$\pm0.3$  & -2.6 & -2.31$\pm0.38$ & -2.0 \\ 
  19400  & -0.8$\pm0.3$  & -1.0 &  --            & -0.7 \\
  22470  & -0.3$\pm0.3$  & -0.1 & -0.38$\pm0.26$ &  0.0 \\
  22920  & -1.5$\pm0.3$  &  --  & -1.32$\pm0.37$ & -1.3 \\
  23408  & -1.1$\pm0.3$  &  --  & -1.54$\pm0.25$ & -1.3 \\
  28843  & -1.1$\pm0.3$  & -0.3 &  0.06$\pm0.24$ &  0.0 \\
  49333  & -0.7$\pm0.3$  &  --  & -0.51$\pm0.32$ & -0.5 \\
  49606  & -1.2$\pm0.3$  &  --  &  --            & -1.0 \\
  51688  & -0.6$\pm0.3$  & -1.0 &  --            & -0.8 \\
  57219  & -2.0$\pm0.3$  &  --  &  --            &  --  \\
  74196  & -0.5$\pm0.3$  &  --  &  --            & -0.4 \\
  125823 & -1.7$\pm0.3$  & -1.4 & -1.18$\pm0.21$ & -1.4 \\
  142301 & -0.7$\pm0.3$  & -0.8 & -0.24$\pm0.37$ & -0.5 \\
  142990 & -1.2$\pm0.3$  &  --  & -0.74$\pm0.27$ &  --  \\
  144334 & -0.4$\pm0.3$  & -0.3 & -0.29$\pm0.28$ & -0.5 \\
  144661 & -0.5$\pm0.3$  &  --  &  --            &  --  \\
  144844 &  0.4$\pm0.3$  &  --  &  --            &  --  \\
  162374 & -1.3$\pm0.3$  &  --  &  --            & -1.4 \\
  175362 & -1.0$\pm0.3$  & -1.0 & -0.4$\pm0.27$  & -0.5 \\
  202671 & -1.3$\pm0.3$  & -1.4 &  --            & -1.1 \\
\hline
\noalign{\smallskip}

\multicolumn{5}{l}{Column 2: this work, Column 3: Glagolevskij (2002)}\\ 
\multicolumn{5}{l}{Column 4: Kochukhov \& Bagnulo (2006), and} \\

\multicolumn{5}{l}{Column 5: G\'omez et al. (1998)}\\
\noalign{\smallskip}
 \hline
\end{tabular}
\end{center}
\end{table}

\subsection{Visual and bolometric absolute magnitudes}
\label{mag-hew}

 Average absolute magnitudes for Ap/Bp stars were obtained by North et al. 
(1997) and M\'egessier (1988a), and for individual objects were reported by  
G\'omez et al. (1998) and Kochukhov \& Bagnulo (2006). Using photometric 
measurements and {\sc hipparcos} astrometric data, G\'omez et al. (1998) 
obtained that the He-weak stellar group has $M_V\!\approx\!-0.2$ mag and that 
the intrinsic dispersion varies from 0.6 to 0.8 mag. All these stars are 
clearly distributed in the Gould Belt and their kinematic behaviour is similar 
to that of the non-peculiar thin disk stars younger than about $10^9$ 
years.\par

\begin{figure}
\centering
\includegraphics[width=8cm,height=8cm,angle=270]{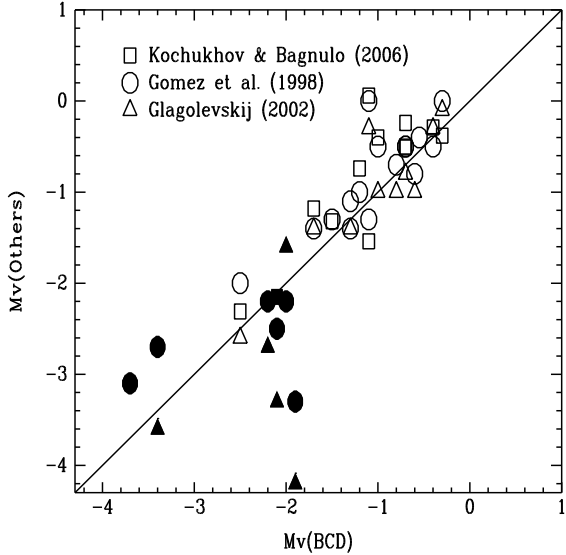}
\caption{Comparison of the visual absolute magnitudes derived from the BCD 
system with those derived by G\'omez et al. (1998), Glagolevskij (2002) and
Kochukhov \& Bagnulo (2006). He-weak stars are represented by open symbols and He-strong stars by filled symbols}
\label{magv}
\end{figure}

 In our sample of 20 He-weak stars, 16 of them have absolute magnitudes 
determined by the LM algorithm (G\'omez et al. \cite{G98}), 10 are 
in common with Glagolevskij (\cite{G02}) and 11 are in common with 
Kochukhov \& Bagnulo's (\cite{KB06}) work (see Table \ref{magni}). 
 Our determinations of visual absolute 
magnitudes are in a very good agreement with those 
obtained by the authors mentioned above (see Figure \ref{magv}). The 
mean discrepancy between the BCD method and the LM algorithm for
the He-weak stars is 0.26 mag. We also 
found a good agreement with $M_{\rm V}$ obtained by Kochukhov \& Bagnulo
(2006), $\Delta M_{\rm V}$ = $\pm$ 0.3. Instead, the mean discrepancy 
between the bolometric magnitudes derived from the BCD system and the
values given by Glagolevskij (\cite{G02}), using the parameter $\beta$ 
of the multicolour photometry, is slightly larger $\sim$  
0.44 mag. This 
discrepancy may be related to the bolometric correction, since when 
transforming the bolometric magnitudes obtained by Glagolevskij 
(\cite{G02}) into visual absolute magnitudes (which are given in Table
\ref{magni}), the discrepancy become smaller.  

Comparing the bolometric absolute magnitudes obtained by us  with the BCD system and the integrated-flux method, we find a mean discrepancy $\Delta\!M_{\rm bol}\!=$ $\!\pm \,$ 0.5 mag. However, a  noticeable disagreement appears for  {HD 144661} and for the He-variable  {HD 125823}. The large difference in the  $M_{\rm bol}$ observed for HD 144661 cannot be recovered by simply changing the color excess $E(B-V)$; by doubling the $E(B-V)$ the bolometric luminosity would increase $\Delta L/L\!\sim\!0.5$, while we need $\Delta L/L\!\sim\!3.4$ to make both estimates agree. Therefore, we may suggest that some undetected flux fading is taking place. This star deserves perhaps further thorough analysis.\par

\section{He-strong stars}

\begin{figure*}[ht]
\centering
\includegraphics[width=15cm,angle=270]{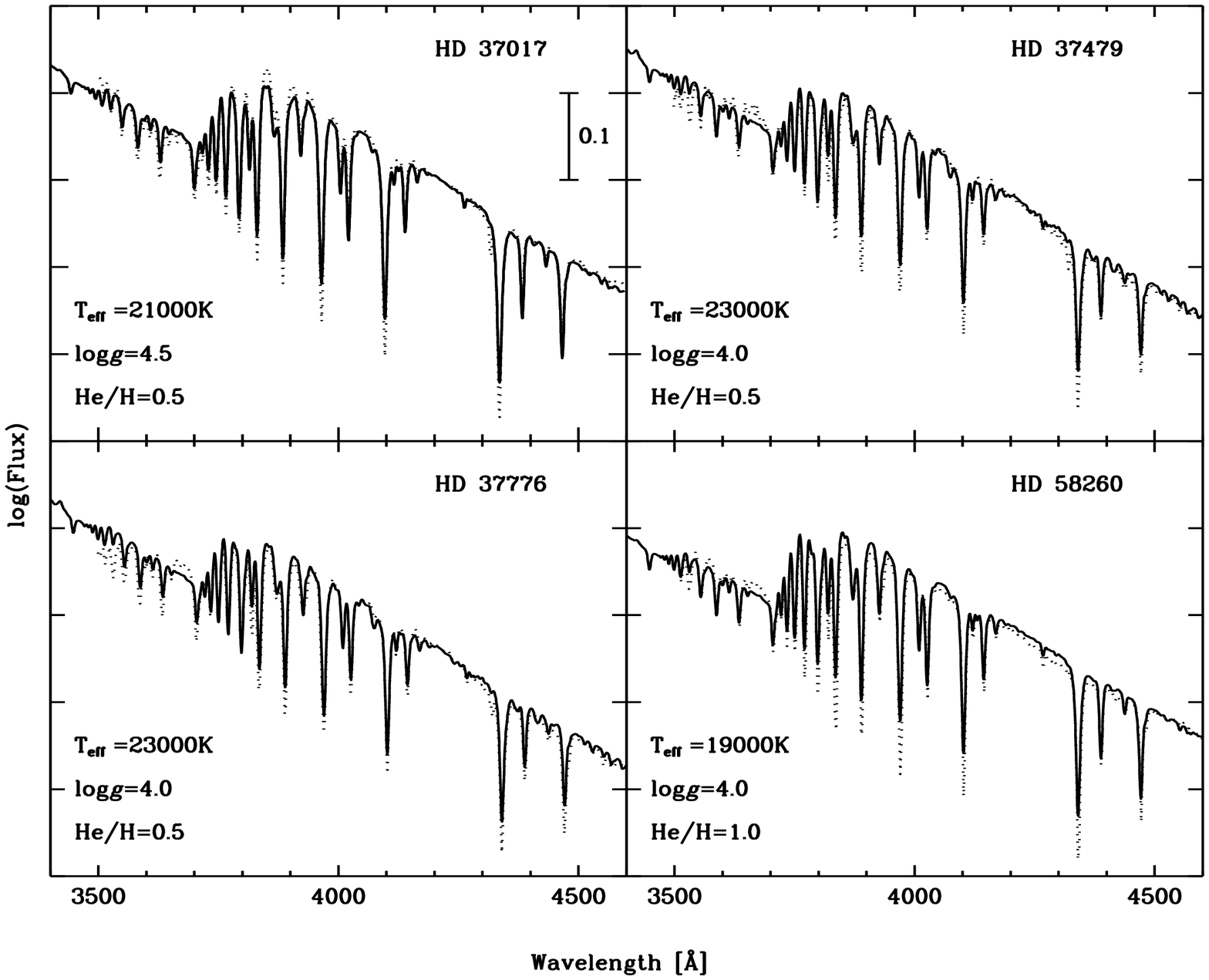}
\caption{Energy distributions of He-strong stars, $\log F_{\lambda}$, around
the BD (dotted line), together with the best theoretical fitting (solid line). 
All spectra have the same flux scale. The convergence of the last He\,{\sc i} 
line members towards 3422 \AA\ clearly indicates the He\,{\sc i} bound-free 
discontinuity}
\label{model1}
\end{figure*}

\begin{figure*}[ht]
\centering
\includegraphics[width=15cm,angle=270]{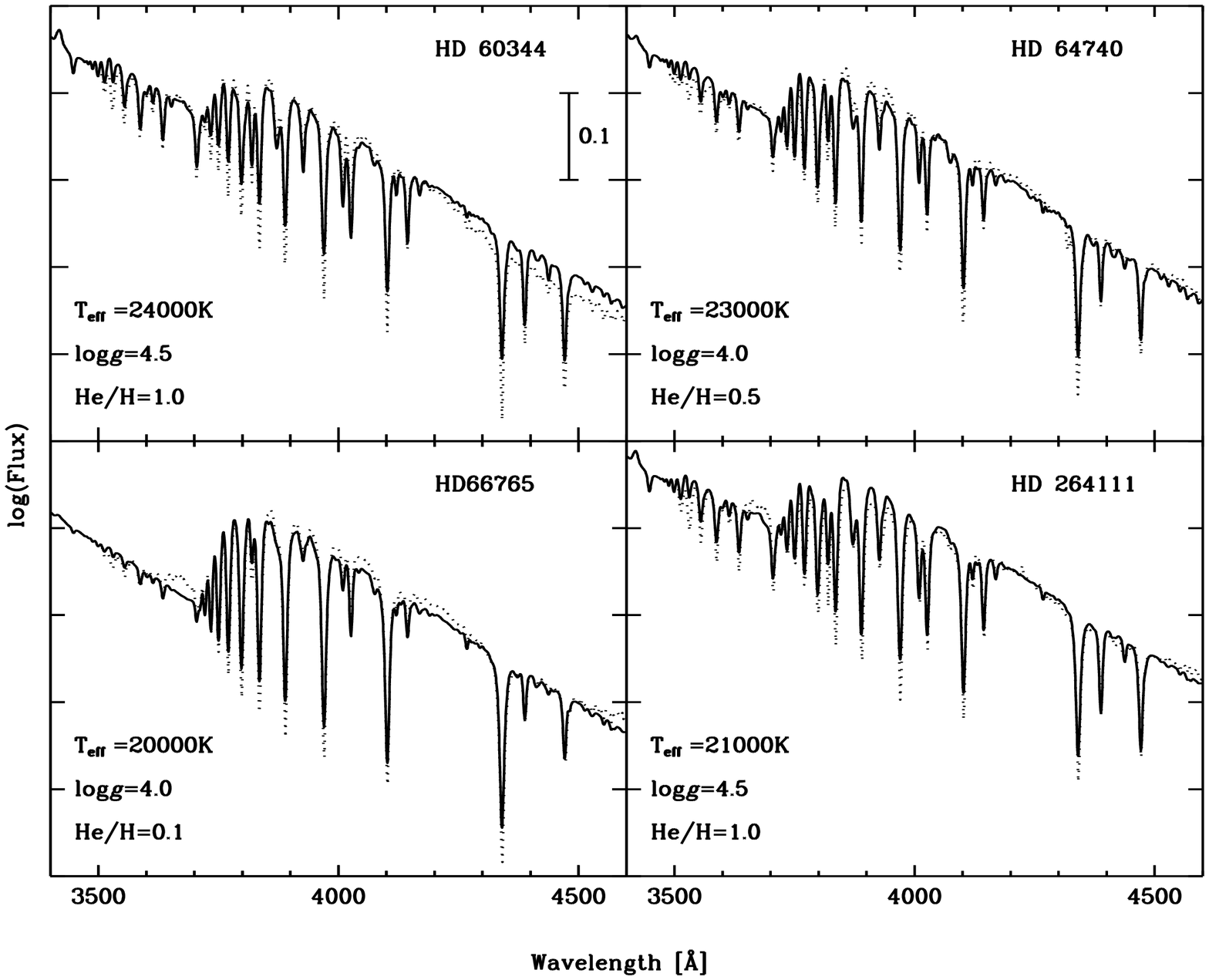}
\caption{Energy distributions of He-strong stars, $\log F_{\lambda}$, around
the BD (dotted line), together with the best theoretical fitting (solid line). 
All spectra have the same flux scale. The convergence of the last He\,{\sc i} 
line members towards 3422 \AA\ clearly indicates the He\,{\sc i} bound-free 
discontinuity}
\label{model2}
\end{figure*}

 The shape of the Balmer continuum of He-strong stars is somewhat different 
from that predicted by models with standard solar helium abundances. The 
He-strong spectra exhibit intense He\,{\sc i} lines and a remarkable 
convergence of the last members of the series to the bound free discontinuity at 3422 \AA, which corresponds to the ionization of the 2--3 Po level (see Figs.~\ref{model1} and \ref{model2}). 
The He-strong stars deserve a special discussion. They may display H$\alpha$ 
emission and spectroscopic and photometric variability. Not only do they show variations in line 
intensities, radial velocities, luminosity and colours but also in the 
magnetic field strength (Renson et al. \cite{R91}).\par
 We have attempted at determining fundamental parameters of these objects by 
means of the BCD system. However, due to the strong He\,{\sc i} lines, a 
precise determination of the Paschen continuum level becomes difficult. 
Nevertheless, the continuum fluxes at $\lambda\lambda$4055 and 4566 \AA\ are unaffected by strong line absorptions (Adelman \& Pyper \cite{AP85}). Thus, we use these points, together with the last members 
of the Balmer line series as references to determine the Balmer jump. 
Table~\ref{he-s} gives our ($\lambda_1,D$) measurements and the corresponding 
fundamental parameters. Data for the He-strong star HD 66765 is scarce and spectrophotometric measurements are published for the first time.\par

\begin{table}[t]
\caption{He-strong stars: measurements and fundamental parameters derived from
the BCD spectrophotometric system}
\label{he-s}
\begin{center}
\tabcolsep 4.0pt
\begin{tabular}{rccrclcc}
\hline\hline
\noalign{\smallskip}
HD & $\lambda_1$ & $D$ & $T_{\rm eff}$ & $\log g$ & \scriptsize{Sp.T} & $M_{\rm V}$ &  $M_{\rm bol}$ \\
     & \AA         & dex & K             & dex      &      & mag         &  mag \\
\hline
\noalign{\smallskip}
{ { 37017}} & 68.72 & 0.145 & 21700 & 4.22 &\scriptsize{B1.5V} & $-$2.0 & $-$4.5 \\
{ { 37479}} & 87.77 & 0.107 & 25300 & 4.30 &\scriptsize{B0.5V} & $-$2.0 & $-$5.3 \\
{ { 37776}} & 74.68 & 0.107 & 25400 & 4.25 &\scriptsize{B0.5V} & $-$2.2 & $-$5.5 \\
{ { 58260}} & 66.05 & 0.149 & 21400 & 4.20 &\scriptsize{B2V}   & $-$1.9 & $-$4.4 \\
{ { 60344}} & 52.97 & 0.111 & 24600 & 3.80 &\scriptsize{B1IV}  & $-$3.4 & $-$5.7 \\
{ { 64740}} & 77.97 & 0.115 & 24500 & 4.23 &\scriptsize{B1V}   & $-$2.1 & $-$5.2 \\
{ { 66765}} & 60.79 & 0.160 & 20200 & 4.11 &\scriptsize{B2V}   & $-$2.0 & $-$4.4 \\
{ {264111}} & 54.78 & 0.092 & 26700 & 3.85 &\scriptsize{B0IV}  & $-$3.7 & $-$6.5 \\
\noalign{\smallskip}
\hline
\noalign{\smallskip}
\noalign{\smallskip}
\multicolumn{8}{l}{Note: The $\lambda_1$ parameters are given in $\lambda-3700$ \AA}\\
\noalign{\smallskip}
\hline 
\end{tabular}
\end{center}
\end{table}

\begin{table}
\begin{center}
\caption{Stellar parameters for the reference He-strong stars obtained with 
NLTE model atmospheres. We list the HD numbers (column 1), the adopted $T_{\rm eff}$, $\log g$, and He/H ratio (columns 2, 3, and 4, respectively), and  $V\!\sin i$ (column 5). $V\!\sin i$ values are taken from Vet\"o et al. (\cite{V91}), Walborn (\cite{W83}), and Zboril et al. (\cite{Z99})}
\label{tab:herich}
\begin{tabular}{lclcr}
\hline
\hline
HD & $T_{\rm eff}$(NLTE) &  $\log g$ & He/H & $V\!\sin i$ \\
      & [K] & [dex]     &        & km~s$^{-1}$ \\
\hline
 37017 & 21000  & 4.50 & 0.5    &   95 \\
 37479 & 23000  & 4.00 & 0.5    &  162 \\
 37776 & 23000  & 4.00 & 0.5    &   95 \\
 58260 & 19000  & 4.00 & 1.0    &   45 \\
 60344 & 24000  & 4.50 & 1.0    &   55 \\
 64740 & 23000  & 4.00 & 0.5    &  160 \\
 66765 & 20000  & 4.00 & 0.1    &  100 \\
264111 & 21000  & 4.50 & 1.0    &  140 \\
\hline
\end{tabular}
\end{center}
\end{table}

 The energy distributions of the observed He-strong stars were fitted with 
 the TLUSTY and SYNSPEC computing codes to have another independent effective temperature estimate. All models were convolved with a rotational broadening function for the 
respective stellar $V\!\sin i$ parameters. In Figures~\ref{model1} and 
\ref{model2} we show the best fit obtained for each studied star. These fits 
are for the ($T_{\rm eff}$, $\log g$, H/He ratio) sets of parameters given in 
Table~\ref{tab:herich}. The uncertainties on the parameters  obtained by model fitting are on average: 1000 K for $T_{\rm eff}$ and 0.5 dex 
in $\log g$. However, for each He-strong star, the model that best 
fits the continuum energy distribution  does not correctly fit the
intensity of  H lines. A better fitting of H lines would require models with 
larger atmospheric temperature gradients than the ones used in NLTE blanketed plane-parallel, in hydrostatic and radiative equilibrium, classical models. Such 
differences in the atmospheric temperature structure could be due not only to
possible non-uniform distributions of the 
abnormally abundant chemical elements in latitude and/or in depth, but also to a non-negligible role of 
the magnetic field in cooling the upper atmospheric layers.\par

\subsection{$T_{\rm eff}$ determination}
\label{TeffHe-s}

 We have sought for previous determinations of effective temperatures in the literature. {Most of the $T_{\rm eff}$ values}  were in works that deal with only  a reduced number of stars  and use different techniques (Hunger \& Groote \cite{HG99}; Leone et al. \cite{L97}; Zboril et al. \cite{Z97}; 
Glagolevskij \cite{G02}; Theodossiou \& Danezis \cite{TD91}; Adelman \& Pyper
\cite{AP85}; Kaufmann \& Theil \cite{KT80}). 

 Table~\ref{compare2} compares the BCD effective temperatures with those obtained by Glagolevskij (\cite{G02}), Zboril et al. (\cite{Z97}) and Kaufmann \& Theil (\cite{KT80}). In this Table, the non-BCD temperature determinations 
are from theoretical LTE model-fitting (Kurucz \cite{kur92}), or from the 
Geneva photometry. We have chosen only these works because they treat a
non negligible number of stars that are in common with our program stars. 
 We found large
discrepancies among the $T_{\rm eff}$ estimated by different authors: 
$2000\!\la\!\Delta T_{\rm eff}\la\!5000$ K. In the same way, our BCD determinations do not escape the rule. Moreover, we observed large differencies between the $T_{\rm eff}$ derived with the BCD method and the NLTE models we used. In most cases, the $T_{\rm eff}$(BCD) are
systematically larger than the average $T_{\rm eff}$ from other methods. This
difference deserves a more detailed discussion that will be done using the stellar atmosphere models in \S \ref{ssynthetic}.\par

\begin{table}[h]
\caption{Comparison of the BCD effective temperatures with those determined by 
other authors for the program He-strong stars}
\label{compare2}
\tabcolsep5.0pt
\begin {tabular}{crccccl}
\hline\hline
\noalign{\smallskip}
& HD  & \multicolumn{5}{c}{$T_{\rm eff} \times10^3$ K} \\
\noalign{\smallskip}
\cline{3-7} 
\noalign{\smallskip}
&    &  BCD & G & Z & K \\ 
\noalign{\smallskip}
\hline
\noalign{\smallskip}
& 37017  & 21.7$\pm1.5$ & 20.10$\pm0.3$  & 19.2  & 21.0 \\
& 37479  & 25.3$\pm1.5$ & 22.45$\pm0.3$  & 22.2  & 23.5 \\
& 37776  & 25.4$\pm1.5$ & 23.35$\pm0.3$  & 21.8  & 26.0 \\
& 58260  & 21.4$\pm1.5$ & 19.70$\pm0.3$  & 19.0  &  --   \\
& 60344  & 24.6$\pm1.5$ & 21.90$\pm0.3$  & 21.7  & 25.0$\pm$1.0\\
& 64740  & 24.5$\pm1.5$ & 23.85$\pm0.3$  & 22.7  & 23.5 \\
& 66765  & 20.2$\pm1.5$ &  --            &  --   &  --   \\
& 264111 & 26.7$\pm1.5$ & 21.60$\pm0.3$  & 23.2  & ~20.0 \\ 
\noalign{\smallskip}
\hline
\noalign{\smallskip}
\multicolumn{7}{l}{Column 2: BCD system; Column 3: } \\
\multicolumn{7}{l}{Glagolevskij (2002); Column 4: Zboril et al. (1997);} \\ 
\multicolumn{7}{l}{Column 5: Kaufmann \&Theil (1980)} \\
\hline
\end{tabular}
\end{table}

\subsection{The surface gravity}

As it is mentioned in \S \ref{slogg}, the $\log\,g$ values found in the literature were derived by different methods. In Table \ref{logghe-s} we show BCD $\log\, g$ values together with the determinations given by other authors. The resulting mean discrepancies in $\log\,g$ are 0.36, 0.21, 0.37 and 0.22 dex, respectively. The values of $\log\,g$ quoted in the atlas of Kaufmann \& Theil (1980) are taken from different sources and their uncertainties are not accessible.  

\begin{table}
\begin{center}
\caption{Comparison of surface gravity estimates for the He-strong stars}
\label{logghe-s}
\tabcolsep 3.0pt
\scriptsize
\begin{tabular}{rrcccc}
\hline
\hline
~~ HD &$\log\!g$(BCD) & $\log\!g$(Z) & $\log\!g$(HG) & $\log\!g$(LCM) & $\log\!g$(K) \\
\hline
\noalign{\smallskip}

  37017  & 4.22$\pm$0.2  & 4.45$\pm$0.10  &  4.12$_{-0.23}^{+0.34}$ &  4.02$\pm$0.25 & 4.4 \\   
  37479  & 4.30$\pm$0.2  & 4.53$\pm$0.17  &  --                     &  3.86$\pm$0.25 & 4.1 \\ 
  37776  & 4.25$\pm$0.2  & 4.52$\pm$0.07  &  4.16$_{-0.30}^{+0.30}$ &  4.11$\pm$0.25 & 4.0 \\      
  58260  & 4.20$\pm$0.2  & 4.02$\pm$0.18  &  3.55$_{-0.27}^{+0.47}$ &  3.52$\pm$0.25 & --  \\
  60344  & 3.80$\pm$0.2  & 4.48$\pm$0.18  &  --                     &  3.50$\pm$0.25 & 3.62 \\  
  64740  & 4.23$\pm$0.2  & 4.50$\pm$0.06  &  4.10$_{-0.08}^{+0.09}$ &  3.80$\pm$0.25 & 3.9 \\ 
  66765  & 4.11$\pm$0.2  &  --            &  --                     &  --            & --  \\
  264111 & 3.85$\pm$0.2  & 4.54$\pm$0.06  &  --                     &  --            & ~4.0 \\
\hline
\noalign{\smallskip}
\multicolumn{6}{l}{Column 2: this work, Column 3: Zboril et al. (1997),}\\ 
\multicolumn{6}{l}{Column 4: Hunger \& Groote (1999), Column 5:}\\
\multicolumn{6}{l}{Leone et al. (1997), Column 6: Kaufmann \& Theil (1980)}\\
\hline
\end{tabular}
\end{center}
\end{table}

The best agreement is obtained with values derived by Hunger \& Groote (1999) from {\sc hipparcos} parallaxes together with tracks from stellar evolution.

\subsection{Visual and bolometric absolute magnitudes}

The He-strong group typically has $M_{\rm V} \!\sim\!-1.6$ mag with a dispersion of about 1.2 mag (G\'omez et al 1998) .\par 

\begin{table}
\begin{center}
\caption{Absolute visual magnitudes for the program He-strong 
stars}
\label{maghe-s}
\tabcolsep 4.0pt
\begin{tabular}{rrccc}
\hline
\hline
~~ HD &$M_{\rm V}$(BCD) & $M_{\rm V}$(G) & $M_{\rm V}$(KB) & $M_{\rm V}$(Go) \\
\hline
\noalign{\smallskip}

  37017  & -2.0$\pm$0.3  & -1.6 &  --   & -2.2 \\ 
  37479  & -2.0$\pm$0.3  &  --  &  --   &  --  \\  
  37776  & -2.2$\pm$0.3  & -2.7 &  --   & -2.2 \\
  58260  & -1.9$\pm$0.3  & -4.2 &  --   & -3.3 \\ 
  60344  & -3.4$\pm$0.3  & -3.6 &  --   & -2.7 \\
  64740  & -2.1$\pm$0.3  & -3.3 & -2.15$\pm$0.25 & -2.5 \\ 
  66765  & -2.0$\pm$0.3  &  --  &  --   &  --  \\
  264111 & -3.7$\pm$0.3  &  --  &  --   & -3.1 \\
\hline
\noalign{\smallskip}
\multicolumn{5}{l}{Columns 2: this work, Column 3: Glagolevskij (2002),}\\ 
\multicolumn{5}{l}{Column 4: Kochukhov \& Bagnulo (2006), and}\\
\multicolumn{5}{l}{Column 5: G\'omez et al. (1998)}\\
\hline
\end{tabular}
\end{center}
\end{table}

 In our sample of 8 He-strong stars, 6 of them have absolute magnitudes 
determined by the LM algorithm (G\'omez et al. \cite{G98}), 5 are 
in common with Glagolevskij (\cite{G02}) and 1 with 
Kochukhov \& Bagnulo's (\cite{KB06}) work (see Table \ref{maghe-s}). In Table \ref{maghe-s}, uncertainties for each value are listed when available. The values corresponding to  $M_{\rm V}$(G) were derived from   $M_{\rm bol}$ given by Glagolevskij (2002) adding the respective bolometric correction.    
 Our determinations of visual magnitudes are in good agreement with those 
obtained by the authors mentioned above (see Figure \ref{magv}, filled symbols), with the exception of HD 58260. We also found that for HD 64740 $M_{\rm V}$(BCD) differs significantly with the value obtained by Glagolevskij (2002).
We find an average discrepancy for all the stars in common of 0.53 mag between the $M_{\rm V}$ obtained with the BCD method and the LM algorithm. Comparing the values of $M_{\rm bol}$(BCD) with the absolute bolometric magnitudes given by Glagolevskij (\cite{G02}) we find an average discrepancy of 0.71 mag. This large difference may probably due to the bolometric correction used by Glagolevskij (\cite{G02}) as we mentioned in \S \ref{mag-hew}.

\subsection{Other relevant results}
\label{ssynthetic}
{\it a) - Synthetic energy distributions as a function of the  He/H abundance ratio}
\medskip

Since BCD determinations of $T_{\rm eff}$ for the He-strong stars are
systematically higher than those obtained by other authors,  we calculate
low-resolution synthetic spectra, in order to quantify the influence of the 
He/H abundance ratio on the emitted visual energy distribution. From these spectra we obtain the BCD parameters following 
the same procedure as for the empirical ones. Models are computed for 
effective temperatures close to those inferred empirically for He-strong stars 
and only for the wavelength interval of the observed low-resolution spectra.\par
\begin{figure}[t]
\centering
\includegraphics[width=7cm,angle=270]{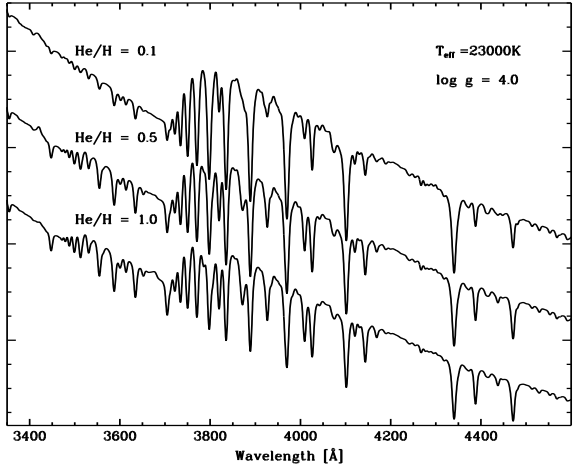}
\caption{Synthetic spectra displaying line and continuum changes according 
to increased He abundances. The spectra are slightly displaced. The NLTE 
model computations were performed for $T_{\rm eff}$ = 23000 K, $\log g$ = 4.0 
dex and ratios H/He= 0.1 (standard), 0.5 and 1.0}
\label{synthetic}
\end{figure}

 An illustration of the effects produced on the visible energy distributions
 resulting from increased abundance ratios He/H is shown in Figure~\ref{synthetic}.
The spectra are slightly displaced in order to see clearly the changes in the
BD, where it is apparent its shortening as He/H is higher. In this figure we 
can also notice the changes on the visible Balmer energy distribution induced 
by enhanced He free-bound transitions and the presence of strong  He\,{\sc i} line 
absorptions. In Table~\ref{models}, the obtained D values are given as a 
function of the model nominal ($T_{\rm eff},\log g$, He/H) parameters. Thus,
increased He/H ratios produce smaller values of D. This reduction has only a 
second order dependence with $\log g$. The lowering $\delta$D of the BD, 
averaged over $\log g$, as function of the effective temperature is given 
within 99.8\% of approximation by the following interpolation relation:

\begin{equation}
\delta{\rm D} = -0.056\left[1-0.233\left(\frac{T_{\rm eff}}{10^4}\right
)^{0.974}\right]({\rm He/H}) \ \ \ \ \ {\rm (dex)}
\label{lowd}
\end{equation}

\noindent Relation (\ref{lowd}) implies that the higher the $T_{\rm eff}$ the
smaller the $\delta{\rm D}$. However, as D $\propto 1.0/T_{\rm eff}$, small 
changes in D carry big differences in $T_{\rm eff}$. Since the BCD 
calibrations used to estimate the effective temperatures of He-strong stars 
are actually made for objects with He/H $\approx0.1$, from Table~\ref{models}
and relation (\ref{lowd}), it becomes clear that the obtained BCD $T_{\rm eff}$ are  overestimated.\par
 Models also predict a slight brightening of stars as the He/H ratio increases. 
Table~\ref{thmag} reproduces the changes of the monochromatic magnitude at
$\lambda4250$ \AA\ with respect to models for He/H = 0.1 that stand for the 
``normal" or standard solar abundance ratio. We notice in Table~\ref{thmag}
that the brightening effect is only marginally dependent on $T_{\rm eff}$ and 
$\log g$.\par

\begin{table}[h]
\caption{Model D parameters as a function of $T_{\rm eff}$,
$\log g$ and He/H abundance ratio}
\label{models}
\begin{center}
\begin {tabular}{cccccc}
\hline\hline
\noalign{\smallskip}
 $T_{\rm eff}$ & $\log g$ & \multicolumn{4}{c}{He/H} \\
\noalign{\smallskip}
\cline{3-6}
\noalign{\smallskip}
    &   & 0.1 & 0.2 & 0.5 & 1.0 \\ 
\noalign{\smallskip}
\cline{3-6}
\noalign{\smallskip}
    &   & \multicolumn{4}{c}{D (dex)} \\ 
\hline
\noalign{\smallskip}
 17000  & 3.5  & 0.206  & 0.201  & 0.189 & 0.170 \\
        & 4.0  & 0.217  & 0.207  & 0.199 & 0.184 \\
        & 4.5  & 0.218  & 0.215  & 0.207 & 0.191 \\
\noalign{\smallskip}
 19000  & 3.5  & 0.176  & 0.169  & 0.156 & 0.140 \\
        & 4.0  & 0.190  & 0.179  & 0.169 & 0.144 \\
        & 4.5  & 0.190  & 0.184  & 0.169 & 0.155 \\
\noalign{\smallskip}
 21000  & 3.5  & 0.135  & 0.133  & 0.122 & 0.110 \\
        & 4.0  & 0.136  & 0.142  & 0.130 & 0.114 \\
        & 4.5  & 0.158  & 0.151  & 0.140 & 0.123 \\
\noalign{\smallskip}
 23000  & 3.5  & 0.111  & 0.110  & 0.104 & 0.098 \\
        & 4.0  & 0.126  & 0.123  & 0.111 & 0.099 \\
        & 4.5  & 0.137  & 0.131  & 0.119 & 0.106 \\
\noalign{\smallskip}
 25000  & 3.5  & 0.092  & 0.089  & 0.080 & 0.069 \\
        & 4.0  & 0.089  & 0.095  & 0.086 & 0.077 \\
        & 4.5  & 0.107  & 0.103  & 0.090 & 0.077 \\
\noalign{\smallskip}
 27000  & 3.5  & 0.080  & 0.077  & 0.071 & 0.063 \\
        & 4.0  & 0.092  & 0.087  & 0.079 & 0.068 \\
        & 4.5  & 0.094  & 0.089  & 0.082 & 0.074 \\
\noalign{\smallskip}
 30000  & 3.5  & 0.059  & 0.057  & 0.050 & 0.042 \\
        & 4.0  & 0.072  & 0.066  & 0.061 & 0.054 \\
        & 4.5  & 0.074  & 0.073  & 0.068 & 0.059 \\
\noalign{\smallskip}
\hline
\end{tabular}
\end{center}
\end{table}
 
We notice that the BCD parameters, $D$, derived from the fitted spectra are 
consistent with those measured in the observed spectra. We can then assert 
that the actual Balmer jump of the studied He-strong stars is well-represented by the empirical $D$ parameter. Then, entering Table~\ref{models} with the measured D values (Table~\ref{he-s}) in columns with high He/H ratios, we recover approximately the effective temperatures obtained by fitting NLTE models (Table~\ref{tab:herich}). These $T_{\rm eff}$(NLTE) values are similar to those reported by other authors (Table~\ref{compare2}).\par

\begin{table}[h]
\caption{Model monochromatic magnitude brightening at $\lambda4250$ \AA\ as 
a function of He/H, $T_{\rm eff}$ and $\log g$}
\label{thmag}
\begin{center}
\begin {tabular}{ccccc}
\hline\hline
\noalign{\smallskip}
 $T_{\rm eff}$ & $\log g$ & \multicolumn{3}{c}{He/H} \\
\noalign{\smallskip}
\cline{3-5}
\noalign{\smallskip}
    &   & 0.2 & 0.5 & 1.0 \\ 
\noalign{\smallskip}
\cline{3-5}
\noalign{\smallskip}
  [K]  & [dex]  & \multicolumn{3}{c}{$\Delta M_{4250}$ [mag]} \\ 
\hline
\noalign{\smallskip}
 17000  & 3.5  & -0.013  & -0.046 & -0.089 \\
        & 4.0  & -0.011  & -0.039 & -0.075 \\
        & 4.5  & -0.009  & -0.031 & -0.061 \\
\noalign{\smallskip}
 19000  & 3.5  & -0.017  & -0.052 & -0.112 \\
        & 4.0  & -0.017  & -0.055 & -0.100 \\
        & 4.5  & -0.010  & -0.047 & -0.087 \\
\noalign{\smallskip}
 21000  & 3.5  & -0.015  & -0.055 & -0.107 \\
        & 4.0  & -0.016  & -0.059 & -0.113 \\
        & 4.5  & -0.016  & -0.058 & -0.111 \\
\noalign{\smallskip}
 23000  & 3.5  & -0.013  & -0.048 & -0.093 \\
        & 4.0  & -0.015  & -0.055 & -0.106 \\
        & 4.5  & -0.016  & -0.058 & -0.112 \\
\noalign{\smallskip}
 25000  & 3.5  & -0.007  & -0.040 & -0.085 \\
        & 4.0  & -0.002  & -0.042 & -0.090 \\
        & 4.5  & -0.007  & -0.050 & -0.100 \\
\noalign{\smallskip}
 27000  & 3.5  & -0.010  & -0.039 & -0.074 \\
        & 4.0  & -0.002  & -0.038 & -0.077 \\
        & 4.5  & -0.015  & -0.051 & -0.094 \\
\noalign{\smallskip}
 30000  & 3.5  & -0.010  & -0.033 & -0.047 \\
        & 4.0  & -0.020  & -0.047 & -0.084 \\
        & 4.5  & -0.023  & -0.047 & -0.087 \\
\noalign{\smallskip}
\hline
\end{tabular}
\end{center}
\end{table}
  
\medskip
{\it b) - Spectral variations}
\medskip

\begin{figure}[ht]
\centering
\includegraphics[width=8.5cm]{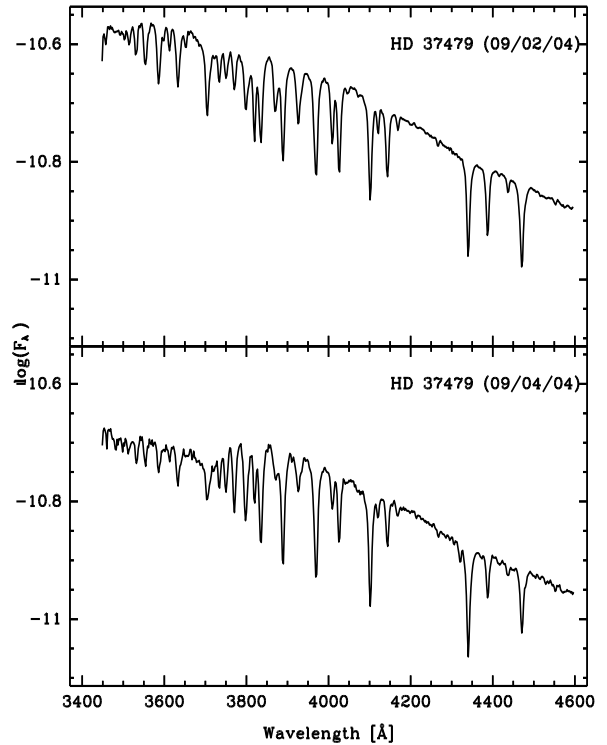}
\caption{Line variations observed in the spectrum of HD 37479. The Balmer  
continuum appears in emission in the spectrum taken in Sep. 2, 2004}
\label{HD37479}
\end{figure}

  We observe that  {HD 37479} (Figure \ref{HD37479}) has near-UV flux, Balmer 
jump and line intensity variations, while the Paschen continuum does not seem 
to undergo appreciable changes. The intensity of H lines increases when that 
of He\,{\sc i} lines decreases and the near-UV flux is lower. A remarkable 
difference is found in the equivalent widths ($\sim$ 30 \%), in the line 
intensities, and on the He/H line ratio as measured in the spectrum obtained
on 2004, Sep. 2 and in another one taken two days before (see Table~\ref{EW}). 
In the spectrum of Sep. 2, there is a flux excess in the Balmer 
continuum near the BD, as compared to its level in the spectrum of two days 
latter. This excess is reminiscent of the second component of the BD seen in
emission in some Be stars (Divan 1978). Whether this flux excess in He-strong 
stars is a matter of an actual emission, still remains to be proved.
Nevertheless, if we estimate $D$ with the criteria explained in 
\S~\ref{ssynthetic}~a), 
we can see that $D$ $= 0.094\pm0.006$ dex on 2004, Sep. 2 and $D$ $=0.107\pm0.004$ dex on 2004, Sep. 4. This implies that the stellar $T_{\rm eff}$ remains 
fairly unchanged when there are strong variations of He and H lines. 
Eventhough, the flux excess near the BD might lead to too high values of  $T_{\rm eff}$ when derived by model fitting.\par  
 {HD 37017} displays a similar behaviour as HD 37479 while HD 64740 does not show appreciable variations for the observing dates.\par  

\section{Discussion}

 The BCD spectrophotometric classification system has shown to be a useful 
tool to determine fundamental parameters for He-weak stars. To 
this end, we can use the calibrations of the BCD 
($\lambda_1,$D) parameters obtained from stars near the Sun. 

It is well-known that some He-weak stars display photometric and spectroscopic variations and this fact might influence the determinations of fundamental parameters. However, taking into account that we have observed many of the program stars only once, it is interesting to stress the good agreement found between the fundamental parameters obtained with the BCD system and with other techniques corresponding to data taken in different dates and probably at different phases. Then we may suggest that, in general, the variations due to chemical inhomogenities do not introduce significant changes on the BD of He-weak stars. However, it would be necessary to study in detail those highly variable objects where temperature fluctuations or variations of the magnetic field might be plaussible.
Contrary to the He-weak group, He-strong stars show large discrepancies among the $T_{\rm eff}$ values found in the literature.  There are several arguments 
that can account for these discrepancies: a) methods based on the Kurucz' 
(\cite{kur79}) line blanketed solar composition models can only lead to an 
approximate flux calibration but, if we do not have a gravity indicator, it 
is impossible to choose the temperature uniquely (Adelman \& Pyper 
\cite{AP85}); b) the use of H$\beta$ and H$\gamma$ equivalent widths combined
with $T_{\rm eff}$ calibrations and/or theoretical Balmer line fittings are 
uncertain since these lines might be filled by emission or be variable; c) 
Geneva colours are affected by the He-overabundance and, furthermore, 
calibration problems may also be present due to the lack of data for hot 
stars. Zboril et al. (\cite{Z97}) have shown that the [U-B] index of the 
Geneva photometric system is affected by the enhanced He/H ratio, which 
affects the b-f and b-b He transitions. Due to these difficulties, the 
mentioned methods lead to somewhat uncertain $T_{\rm eff}$ determinations.\par

In this work we have shown that the BCD calibrations lead to overestimated values of $T_{\rm eff}$ for the He-strong stars, since they are  actually made for objects with He/H $\approx0.1$. However,  if care is made at identifying the genuine continuum flux  levels, as indicated in \S \ref{ssynthetic}, we can still measure the BCD parameter D. This parameter not only well-represents the actual Balmer jump of  He-strong stars but also reflects closely the true effective temperature of the star and their H/He ratio when it is used in combination with results of NLTE models. An alternative way to determine the effective temperature of He-strong stars could be the use of the integrated flux method, where the determination of the angular diameter $\theta$ is almost insensitive to the type, or quality, of model atmosphere used in the near-IR region. The difference between $T_{\rm eff}$(BCD) and $T_{\rm eff}$(IF) translated into BD difference $\delta D$ may then produce an estimate of the He/H ratio in these objects.\par

For the He-strong stars, the Balmer jump might be affected by changes in He/H ratios rather than by changes of the photospheric temperature. When increasing the abundance of He, stellar atmospheres of B stars become less opaque due
to the transparency of He\,{\sc i} in the visible continuum and the reduction
of the relative number of H absorbing atoms. Direct calculations of the
$\kappa^+/\kappa^-$, total continuum absorption coefficient ratio at the Balmer
discontinuity,  show that this effect is slightly stronger for $\kappa^-$ than
for $\kappa^+$. This leads to smaller values of $D$ for He-strong stars than for stars with standard He abundance. However, it can still be noted that the He abundance increase may concern
only a fraction $f$ of the apparent stellar hemisphere. The fit of observed energy distributions with models should then use both He/H and $f$ as free parameters. The presence of helium patches could be explained in terms of the so-called Oblique Rotator Model. The enhancement of He/H ratio might also give rise to a flux excess in the Balmer continuum originating a second component of the BD in emission. 

\subsection{Evolutionary status of the studied stars}

\begin{figure}
\centering
 \includegraphics[width=6cm,angle=270]{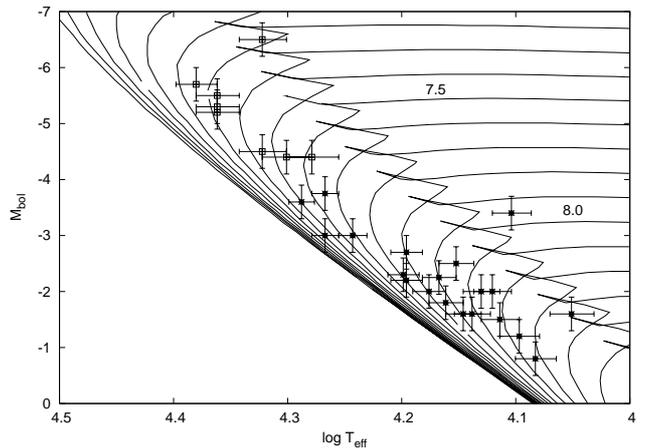}
\caption{HR-diagram for He-weak stars (dots) according to BCD fundamental parameters. We also include the He-strong objects (squares) taking into account, in this case, the $T_{\rm eff}$ derived from NLTE models. Isochrones are from Bressan et al.(\cite{bre1993}). The ages range from 10$^{6.6}$ to 10$^{10}$ yr with steps of $\Delta\log t\! =\!0.1$.}
\label{HR}
\end{figure}

 In order to obtain an insight on the evolutionary status of the He-weak and 
He-strong stars studied in this work, we plot in Figure ~\ref{HR} $M_{\rm bol}$ 
vs. $\log T_{\rm eff}$. For He-strong stars we have taken into account the 
$T_{\rm eff}$ values derived with NLTE models. The isochrones are for normal stars of metal content typical for the solar vicinity, Z=0.02, and were calculated by Bressan et al. (\cite{bre1993}). Some ages  $t$(yr), in logarithm scale, 
are shown in the figure and  the curves are spaced by steps of 0.1 dex.\par
There is a clear distinction made by the effective temperature, which
separates the He-strong from He-weak stars. However, most He-weak and He-strong objects are in Main Sequence phase. \par

Since in  Figure ~\ref{HR} there is a well-marked separation between He-weak and He-strong stars according to their effective temperatures, we are tempted to conclude that stars do not evolve from He-weak to He-strong, or viceversa. The separation between these stellar classes seems to be due mainly to mass-effective temperature related phenomena. In fact, the known He-variable star HD 125823, which is located between the two He-peculiar groups, shows alternatively He-w/He-s features. This  object could provide important clues to disentangle the evolutionary status of the He-peculiar stars.\par 

 Our sample of objects is too small to derive correlations between the star magnetic field and its fundamental parameters. However our determinations of fundamental parameters and  ages are in very good agreement with those estimated by Kochukhov and Bagnulo (\cite{KB06}). These authors find that the surface magnetic flux of CP stars increases with stellar age and mass, and correlates with the rotation period.

\section{Conclusions}

 Several reviews and papers have highlighted the difficulties of obtaining 
the fundamental parameters of CP stars. Often, the determination of effective
temperatures and surface gravities of CP stars by means of photometry and 
model fittings are  more complex than for normal stars.\par
 The main purpose of this work is to estimate the fundamental parameters of 28 
helium-peculiar stars using the BCD spectrophotometric system. The advantage 
of the BCD system is that it provides simultaneously spectral types, $T_{\rm 
eff}$, $\log g$ and bolometric and absolute magnitudes. All these quantities
are obtained from only two parameters drawn from low-resolution spectra which 
are not affected by ISM and circumstellar absorptions, and the techniques at 
obtaining them avoid the spectral regions that are strongly perturbed by 
chemical anomalies of He-peculiar stars. Besides, the effective temperature 
and surface gravity thus obtained are related to deeper atmospheric layers than those where, on average, the spectral lines are formed, since the BCD quantities are obtained from the visible continuum spectrum.\par
 We confirm that the BCD spectrophotometric system gives reliable effective 
temperatures, surface gravities and absolute magnitudes for He-weak stars. Not only are $T_{\rm 
eff}$ values in very good agreement with those obtained with the UBV, multicolour and
Geneva photometry, but also with the IRFM and LTE line-blanketed models. Surface gravities show an excellent agreement with values derived from {\sc
hipparcos} parallaxes and stellar evolutionary tracks. The absolute magnitudes also agree with the values obtained from the {\sc
hipparcos} parallaxes and the mean discrepancy between 
both methods is on average $\pm$ 0.3 mag.\par
 Particular attention is payed in this paper to the effective temperature of 
He-strong stars. The He-strong stars are He-variable and this variability not 
only affects the intensity of He and H lines, but also the apparent 
distribution of continuum fluxes. These changes can originate difficulties 
of determining the effective temperatures of He-strong stars that were 
discussed in this work. The determination of the BCD parameter D, which 
quantifies the Balmer jump, avoids these inconveniences. Nevertheless, the 
parameter $D$ is a function of the He/H abundance ratio. To specify this 
dependence, we have calculated non-LTE model atmospheres. We have obtained 
thus a relation $D$ = $D$(He/H) and could determine the effective temperature of the studied He-strong stars by model fitting. The results show that the higher the He/H ratio, the smaller $D$. From these models it is also apparent that 
increased values of He/H are accompanied by brighter fluxes in the Paschen 
continuum. We have also found that the model fitting produces much lower 
effective temperatures than expected from the $T_{\rm eff}$ = $T_{\rm 
eff}(\lambda_1,$D) BCD calibrations, which actually suit to stars with 
abundance ratios He/H $\approx0.1$. We have then proposed a method of 
determining the He/H of He-strong stars, which is based on the mentioned D = 
D(He/H) relation and on two distinct $T_{\rm eff}$ determinations, one coming
from the up to now used ($\lambda_1,$D) calibration, and the other, which 
could be the integrated-flux method.
 
 We have observed the He-strong star HD 37479, which underwent a rather 
strong variation in the visible spectrum in an interval of 2 days. Although the
fitting  procedure of classical stellar atmosphere models with the observed energy distributions would suggest an apparent variation of the effective temperature, a careful determination of the $D$ parameter shows that this quantity does not vary significantly. We note that in the BCD system $D$ can be constant over a large range of $\lambda_1$ parameters, which may reflect the fact that the observed spectrophotometric variations correspond to changes in the structure of  possible exophotospheric layers, as in Be stars. In fact, Be stars can sometimes exhibit spectrophotometric changes implying constancy of the total absolute magnitude $M_{\rm V}$ (Moujtahid et al. 1998).
 
 Since the brightening of the Paschen continuum is on average not stronger
than some $\Delta M_{4250}\approx$ $-0.10$ mag, the BCD calibration can still 
be a useful tool to obtain a rough estimate of the absolute magnitudes of 
He-strong stars.

  A first insight on the evolutionary status of He-weak and He-strong stars
was obtained. Both types of He-peculiar stars seem to be in the Main Sequence
evolutionary phase. The He-strong stars are situated roughly in the
$T_{\rm eff}\!\ga$ 19000 K region of the HR diagram,  while the He-weak are in the $T_{\rm eff}\!\la$ 19000 K zone of
 the HR diagram for B-type stars. More precise M$_{\rm bol}$ determinations
for a large number of stars are needed to give a deeper insight on the evolutionary status.\par
 Due to the variability observed in the He-strong group, equivalent widths 
of hydrogen and helium lines are given. Data for the He-strong star HD 66765 
are reported for the first time. Although we have obtained a ``normal" He/H 
ratio for this star, we should consider that this object was reported as a 
variable He-strong star by Wiegert \& Garrison (\cite{WG98}).\par


\bigskip

{\large \bf Appendix A Line equivalent widths of the He-strong stars}
\medskip

 Having in mind the strong variations observed in HD 37479 and HD 37017, we considered 
useful to report line equivalent width measurements for all the program 
He-strong stars.

\begin{table}[h]
\caption{Equivalent widths in \AA\ of hydrogen and helium lines for the 
program He-strong stars and the Helocentric Julian Date}
\label{EW}
\tabcolsep 4.5pt
\begin{tabular}{rcccccc}
\hline\hline
\noalign{\smallskip}
 HD    & HJD  &   H$\gamma$  & \multicolumn{4}{c}{He\,{\sc i}} \\
\noalign{\smallskip}
\cline{4-7}
\noalign{\smallskip}
& -2453000 &  & $\lambda$4026 & $\lambda$4144 & $\lambda$4387 & 
$\lambda$4471 \\
\noalign{\smallskip}
 \hline
\noalign{\smallskip}
 37017 &  249.9020  & 5.48  & 2.05  & 1.28 & 1.37 & 2.19 \\
 37017 &  278.7787  & 7.26  & 1.17  & 0.74 & 0.76 & 1.43 \\
 37479 &  250.8538  & 3.20  & 3.08  & 2.10 & 2.12 & 3.32 \\
 37479 &  252.8841  & 4.16  & 2.13  & 1.30 & 1.35 & 2.32 \\ 
 37776 &  252.8756  & 4.00  & 2.29  & 1.52 & 1.42 & 2.53 \\ 
 58260 &  754.7432  & 4.28  & 2.12  & 1.34 & 1.30 & 2.33 \\ 
 60344 &  755.7009  & 4.16  & 1.95  & 1.44 & 1.35 & 2.42 \\ 
 64740 &  251.9172  & 4.20  & 1.97  & 1.14 & 1.13 & 2.08 \\ 
 64740 &  278.8559  & 4.50  & 1.63  & 1.33 & 1.29 & 2.18 \\ 
 66765 &  755.7666  & 5.78  & 1.39  & 0.78 & 0.91 & 1.81 \\
125823 &  754.8733  & 6.25  & 0.46  & 0.23 & 0.22 & 0.51 \\ 
264111 &  755.6925  & 4.29  & 2.13  & 1.63 & 1.51 & 2.45 \\
\noalign{\smallskip}
\hline
\end{tabular}
\end{table}

 We measured equivalent widths (EW) of hydrogen and helium 
lines in the spectra of the 8 observed He-strong stars and of the variable 
He-w/He-s star HD 125823 (see Table~\ref{EW}). We have compared our equivalent 
width measurements with those found in the literature. From this comparison it 
comes out that the EWs derived by Walborn (\cite{W83}) are, in most cases, 
larger than ours by factors ranging from 1.3 to 2, except for the He\,{\sc i} lines observed in HD 37479 (Sep. 4, 2004) and HD 264111. The discrepancies between  Zboril's et al. (\cite{Z97}) EWs and ours are about 10 to 15 \%. The largest 
difference is found for HD 37479 (Sep. 2, 2004), our EWs are larger by a 
factor of 1.4.\par 

\bigskip

\bigskip

{\bf Acknowledgements}: This work was partially supported by the Agencia de Promoci\'on Cient\'{\i}fica y Tecnol\'ogica (BID 1728 OC/AR PICT 
03-12720) and the Programa de Incentivos G11/073 of the National University of La Plata. Y. F. acknowledges financial support from the Belgian Federal Science Policy (projects IAP P5/36 and MO/33/018). This work has made use of CDS data base.  We would like to thank our referee Dr. Mathys for his numerous comments and suggestions which considerably helped to improve our manuscript.


\begin{thebibliography}{}
\bibitem{AP85} 
Adelman, S. J., \& Pyper, D. M. 1985, A\&AS, 62, 279
\bibitem{BC41}
Barbier, D., \& Chalonge, D. 1941, Ann. Astrophys., 4, 30 
\bibitem{BS77} 
Blackwell, D. E., \& Shallis, M. J. 1977, MNRAS, 180, 177
\bibitem {bre1993}
Bressan A., Fagotto F., Bertelli G., \& Chiosi C. 1993, A\&AS, 100, 
647
\bibitem{bbm03}
Bychkov, V. D., Bychkova, L. V., \& Madej, J., 2003, A\&A, 407, 631
\bibitem{CHF92}
Cayrel de Strobel, G., Hauck, B., Francois, P., et al. 1992, A\&AS 95, 273
\bibitem{CD52}
Chalonge, D., \& Divan, L. 1952, Ann. Astrophys., 15, 201
\bibitem{CD73} 
Chalonge, D., \& Divan, L. 1973,A\&A, 23, 69
\bibitem{CD77} 
Chalonge, D., \& Divan, L. 1977, A\&A, 55, 117
\bibitem{CZ01} 
Cidale, L. S., Zorec J., \& Tringaniello L. 2001, A\&A, 368,160
\bibitem{DS84}
Dimitrijevi\'c, M. S., \& Sahal-Brechot, S. 1984, JQSRT, 31, 301
\bibitem{DS90}
Dimitrijevi\'c, M. S., \& Sahal-Brechot, S. 1990, A\&AS, 82, 519
\bibitem{D79}
Divan, L. 1979, in ``Spectral Classification of the Future'', IAU 
Coloquium 47, eds. Mc Carthy, Philip, Coyne, Vatican Observatory, p. 247
\bibitem{dz82} 
Divan, L., \& Zorec, J. 1982, ESA-SP 177, 101 
\bibitem{F77} 
Flower, P. J. 1977, A\&A, 54, 31
\bibitem{FC78}
Freudenstein, S. A., \& Cooper, J. 1978, ApJ, 224, 1079.
\bibitem{GHM74} 
Gerbaldi, M., Hauck, B., \& Morguleff, N. 1974, A\&A, 30, 105
\bibitem{G02}
Glagolevskij, Y. V. 2002, Bull. Spec. Astrophys. Obs., 53, 33
\bibitem{G72} 
Golay, M. 1972, Vistas in Astronomy, 14, 13
\bibitem{G98} 
G\'omez, A. E., Luri, X., Grenier, S., et al. 1998, A\&A, 336, 953 
\bibitem{HN93}
Hauck, B., \& North, P. 1993, A\&A, 269, 403
\bibitem{HN82}
Hauck, B., \& North, P. 1982, A\&A, 114, 23
\bibitem{HL95}
Hubeny, I., \&  Lanz, T. 1995, ApJ, 439, 875
\bibitem{H04} 
Hubrig, S., North, P., Mathys, G. 2000, ApJ, 539, 352
\bibitem{HG99} 
Hunger, K., \& Groote, D. 1999, A\&A, 351, 554
\bibitem{JJ87a} Jaschek, C., \& Jaschek, M. 1987a, "The 
Classification of Stars", Cambridge University Press, p.173
\bibitem{JJ87b} 
Jaschek, C., \& Jaschek, M. 1987b, A\&A, 171, 380
\bibitem{JM75} Johnson, H.L., \& Mitchell, R.I. 1975, Rev. 
Mexicana Astron. Astrof. 1, 299
\bibitem{KT80}
Kaufmann, J. P., \& Theil, U. 1980, A\&As, 41, 271
\bibitem{KB06} 
Kochukhov, O., \& Bagnulo, S. 2006, A\&A, 450, 763
\bibitem{KR87} 
Kroll R., 1987, A\&A, 181, 315 
\bibitem{kur79} 
Kurucz, R. 1979, ApJS, 40, 1 
\bibitem{kur92}
Kurucz, R. 1992, CD-ROM No. 19, 20, 21, Cambridge Mass.: Smithsonian
Astrophysisical Obsrevatoryx
\bibitem{L85} 
Lanz, T. 1985, A\&A, 144, 191
\bibitem{LMB94} 
LeBlanc, F., Michaud, G., \& Babel, J. 1994, ApJ, 431, 388
\bibitem{L74} 
Leckrone, D. S., Fowler, J. W., \& Adelman, S. J. 1974, A\&A, 32, 237
\bibitem{LM97} 
Leone, F., \& Manfr\`e, M. 1997, A\&A, 320, 257
\bibitem{L97} 
Leone, F., Catalano, F. A., \& Malaroda, S. 1997, A\&A, 325, 1125 
\bibitem{M88a} 
M\'egessier, C. 1988a, A\&A, 206, 74
\bibitem{M88b} 
M\'egessier, C. 1988b, A\&AS, 72,551
\bibitem{MD85} 
Moon, T. T., \& Dworetsky, M. M. 1985, MNRAS, 217, 305
\bibitem{Mouj98}
Moujtahid, A., Zorec, J., Hubert, A. M. et al. 1998, A\&AS, 129, 289
\bibitem{NSW93} 
Napiwotzki, R., Sch\"onberner, D., \& Wenske, V. 1993, A\&A, 268, 653
\bibitem{N97}
North, P., Jaschek, C., Hauck, B., et al. 1997, Proceedings of the ESA Symposium `Hipparcos - Venice '97', Venice, Italy, ESA SP-402 (July 1997), p. 239-244
\bibitem{OP74} 
Osmer, P. S., \& Peterson, D. M  1974, ApJ, 187, 117
\bibitem{P99} 
Proffitt, C. R., Brage, T., Leckrone, D., et al. 1999, ApJ, 512, 942 
\bibitem{R91}
Renson, P., Gerbaldi, M., \& Catalano, F. A. 1991, A\&AS, 89, 429
\bibitem{SB87} 
Shore, S. N., \& Brown, D. N. 1987, A\&A, 184, 219
\bibitem{S98} 
Sokolov, N. A. 1998, A\&AS, 130, 215
\bibitem{S95} 
Sokolov, N. A. 1995, A\&AS, 110, 553
\bibitem{SD89} 
Stepie\'n, K., \& Dominiczak, R. 1989, A\&A, 219, 197
\bibitem{TD91} 
Theodossiou, E., \&  Danezis, E. 1991, A\&AS, 183, 91
\bibitem{V91}
Vet\"o, B., Hempelmann, A., Sch\"oneich, W. \& Stahlberg, J. 1991, Astron. Nachr., 312, 133
\bibitem{W83}
Walborn, N. 1983, ApJ, 268, 195
\bibitem{WG98}
Wiegert, P., \& Garrison, R. F. 1998, JRASC, 92, 134
\bibitem{Z97}
Zboril, M., North, P., Glagolevskij, Y. V., \& Betrix, F. 1997, A\&A, 324, 949
\bibitem{Z99}
Zboril, M. \& North, P. 1999, A\&A, 345, 244
\bibitem{ZFC05} 
Zorec, J., Fr\'emat, Y., \& Cidale, L. S., 2005, A\&A, 441, 235
\bibitem{z86} 
Zorec, J. 1986, PhD. Thesis, Universit\'e de Paris, France
\bibitem{ZB91} 
Zorec, J., \& Briot, D. 1991, A\&A, 245, 150
\end{thebibliography}
\end{document}